\documentclass[prb,aps,amsfonts,amssymb,floatfix,showpacs,superscriptaddress]{revtex4}
\usepackage{graphicx}
\usepackage{dcolumn}
\usepackage{bm}
\usepackage{color} 
\usepackage{epsfig}
\usepackage{epstopdf}
\usepackage{amsmath}
\begin{document} 
\title{
Mott-Slater Transition in a Textured Cuprate Antiferromagnet}
\author{R.S. Markiewicz and A. Bansil}
\affiliation{ Physics Department, Northeastern University, Boston MA 02115, USA}
\begin{abstract}
We generalize the concept of vortex phase in a type II superconductor to textured phases, where certain phases can persist over an extended range of perturbations by confining competing phases on topological defects (the vortices in a superconductor).  We apply this model to the pseudogap phase in cuprates, where the relevant topological defects are the antiphase domain walls of an underlying antiferromagnetic (AFM) order.  We demonstrate that this model can describe many key features of intertwined orders in cuprates, and most importantly provide the first clear evidence for the Mott-Slater transition in cuprates.
\end{abstract} 
\maketitle

\section{Introduction}
Unraveling the nature of the pseudogap in cuprates has been a major goal in the field for many years, with implications for the mechanism of high-$T_c$ superconductivity.  A related hope is that unraveling the physics of the cuprate pseudogap can shed light on similar phenomena found in many other correlated materials, including pnictides, nickelates, organic, and heavy fermion compounds.  We have recently demonstrated that several key features of the pseudogap -- including Fermi arcs and a logarithmically-diverging heat capacity at the pseudogap collapse, accompanied by the anomalous growth of the Hall number\cite{Taill1} -- can be understood in terms of the termination of an antiferromagnetic (AFM) phase.\cite{Paper1}  However, this is only half of the story.  A key question is why does the pseudogap not develop long-range order, and, in a similar vein, what is the origin of the intertwined orders\cite{IT,ves} seen in different parts of the pseudogap phase diagram.  Here we assemble various lines of evidence to show that the data can be understood in terms of textures of the AFM phase.

To better understand the picture we have in mind, it is convenient to recall a more familiar analogue.  Vortices are topological defects of a superconductor, stabilized by a circulating flow of charge, but at the cost of suppressing superconductivity in the vortex core.  While metastable in a pure superconductor, they play an essential role when a perturbation is applied which opposes superconductivity.  For a type 2 superconductor in a magnetic field, each vortex traps a quantum of magnetic field, allowing superconductivity to persist to much higher fields than otherwise possible.

Similarly, in undoped cuprates at low temperatures, the AFM order has a small Ising anisotropy, so the topological defects are antiphase boundaries of the AFM order, forming domain walls.  Metastable at zero doping, they can trap doped charges and preserve AFM order to much higher doping than would otherwise be possible, in the form of stripe phases\cite{Tranq1}.  In one cuprate, La$_{2-x}$ Ba$_x$CuO$_4$, this stripe phase is strong enough to quench superconductivity, but in other cuprates, evidence for more fluctuating stripes has been found, but only at low doping, whereas at higher doping they are replaced by charge-density waves (CDWs).  For this reason, stripes are usually considered as secondary manifestations, and not as essential to the pseudogap phase. 

However, we here demonstrate that the stripe-to-CDW transition is a (non-Landau) transition of the dressed defects in this intermediate phase, a manifestation of the Mott-Slater transition\cite{MBMB}.  Whereas in the Mott phase the stripes are insensitive to the Fermi surface, in the Slater phase they become dominated by Fermi surface nesting.  Moreover, the Fermi surface being nested is not the nonmagnetic Fermi surface, but the AFM Fermi surface which plays such a large role in paper I\cite{Paper1}.

The theory of the cuprates is difficult because one must first develop the theory of the homogeneous phase diagram, and then incorporate the effects of the heterogeneous textures.  In paper I\cite{Paper1} we set the stage for this analysis, postulating that the pseudogap represents a phase of mostly short-range antiferromagnetic (AFM) order, dominated by dressed topological defects, specifically charged domain walls.  We showed that, near the end of the pseudogap doping range at hole doping $x=x^*$, where $x^*$ is the doping of pseudogap collapse, the domain walls play a subsidiary role and can be neglected, so a simple theory of AFM collapse can describe the main experimental findings.  As an aside, we note that to explain angle-resolved photoemission (ARPES) data, we needed to include an extrinsic effect, the role of dopant clustering in forming local patches of varying electronic doping.\cite{patches,JennyO}  In the present paper we extend these results, focusing on the intrinsic heterogeneity of nanoscale phase separation (NPS) and stripe phases.

This paper is organized as follows.  In Section II, we review earlier work, focusing on two key issues: (1) the role of the Mott-Slater transition in the homogeneous phase diagram; and (2) how the generic idea of NPS, which should arise in many correlated materials, is exemplified in cuprates by a model of electronic textures growing on topological defects of an underlying AFM order, and how this is captured in {\it ab initio} calculations.  In Section III, we tackle the problem of the stripe-CDW crossover, showing (1) that this crossover represents the experimental signature of the Mott-Slater transition; but (2) that an extra complication arises in that when the sample is cooled through the pseudogap transition, it falls out of equillibrium, leaving excess domain walls behind (Kibble-Zurek physics), which can be monitored in STM studies.  In Section IV we demonstrate that the experimental data are consistect with the new paradigm.  A key finding is that the nesting vector of the CDW-like stripe branch is controlled by hot-spot nesting of the {\it AFM Fermi surface}, thereby completing our proof that these are textures of the underlying AFM phase.  Section V is for Discussion, summarizing the key features of our model, while Section VI presents our Conclusions.

\section{Background}

\subsection{Homogeneous model: Mode-coupling and the Mott-Slater transition}
Our many-body theory demonstrates that the cuprates are dominated by short-range AFM order\cite{AIP}, where the NPS was assumed to be responsible for the lack of long-range order, suggesting a relation between the NPS and domain  walls.  A recent study on the role of vertex corrections in cuprates, specifically those associated with mode coupling\cite{MBMB}, sheds much light on this issue.  It is found that mode coupling can frustrate long-range order in two ways, first via enhanced fluctuations in lower dimensional systems (i.e., Mermin-Wagner physics\cite{MW}), but also by competition of many modes trying to soften at the same time.  This latter result is closely related to McMillan’s notion that strongly correlated charge density wave materials are associated with large bosonic entropy, due to many competing phases\cite{McMill}.  Indeed, it has been shown that mode-coupling calculations capture these effects in electron-phonon coupling systems.\cite{Moti}

For present purposes, the chief result of these studies\cite{RM70,MBMB} is that mode coupling can explain many aspects of Mott physics in real materials.  Thus, any magnetic order in half filled cuprates has energy per copper $-U/2$ up to order $J=4t^2/U$, so that the conventional random phase approximation (RPA) breaks down.  In RPA, one simply takes the order parameter yielding the lowest energy and ignores the others, leading in cuprates to a transition to the correct ground state, but starting at $T\sim U$.  Instead the mode coupling model accounts for the entropy of competing modes.  The net result is that a gap opens at $T\sim U$ while a particular order only starts to grow at $T\sim J$.  Thus, in contrast to this failure of the RPA, the mode coupling model properly captures the large reduction of the Neel temperature $T_{N}$.

Beyond explaining the opening of a gap without an accompanying phase transition, the model describes other aspects of Mott physics.  For example, Fermi surface nesting is absent.  Moreover, there is a non-Landau type phase transition from this Mott phase to a Slater phase where such nesting is present.  The transition can be triggered by either doping or by increasing the magnitude of the hopping parameter $t'$. 
Exactly at the transition the correlation length $\xi$ of the AFM order collapses, leading to an emergent spin liquid.  On the Slater side the correlation length recovers but is much smaller -- an order of magnitude smaller at half filling.\cite{MBMB}.  Similar but weaker results have been found in a model of a three-dimensional AFM.\cite{Kohn}  The transition is fairly insensitive to the hopping parameters $t'$, $t''$, so that in LSCO it occurs at a doping $x_{MS}>x_{VHS}$, while for other cuprates $x_{MS}<x_{VHS}$.  Thus, it should be observable in most cuprates as a sort of {\it two-stage pseudogap collapse} as doping is varied.  

 In a uniform system, this Mott-Slater (MS) transition would result in a transition vs doping from an AFM phase which is optimal at $(\pi,\pi)$ to a CDW whose $Q$-vector is controlled by Fermi surface nesting.  Below we shall show how this behavior is reflected in the cuprates, where NPS is present, as a crossover in the nature of the topological defects.  In the Mott phase, there is no Fermi surface nesting, so the stripe $q$-vector is controlled solely by repulsive forces -- the stripes stay as far apart as possible, as found in DFT studies of undoped YBCO$_6$\cite{YUBOI}.  In contrast, YBCO$_7$ has a minimum at a finite $q$-value\cite{YUBOI}, as expected in the Slater phase.  Consistent with this, STM studies find a commensurate ($P_c=4$) to incommensurate ($P_c>4$) phase transition in Bi2201 near $x=0.14$.\cite{EHud2} We will demonstrate that this incommensurate phase is controlled by Fermi surface nesting, and hence is a signature of the Mott-Slater transition.  In Section III, we expand on the nature of the resulting CDW-like phase.

 \subsection{Heterogeneous models}
\subsubsection{Nanoscale phase separation}
We begin by reviewing earlier ideas on NPS, mainly as applied to cuprates.  One of the earliest classes of strongly correlated materials was electron-hole droplets in photoexcited semiconductors, the result of a gas-liquid transition of excitons at high excitation densities.\cite{Keldysh}  The question arises: are holes necessary, or could an electron gas undergo a gas-liquid transition?  The terminology Fermi gas and Fermi liquid suggests that this is possible, but there is a complication. When equal numbers of electrons and holes are present, or if the dopant ions are mobile, a liquid droplet can form since the system remains charge neutral.  On the other hand, the electrons cannot all bunch up by themselves, due to long-range Coulomb interaction.  A solution follows from the electrostatic version of Saint Venant's principle\cite{StVen}: if the electronic phase separation is on a sufficiently fine length scale that no dipole moment builds up, then the Coulomb energy remains small, and a NPS can arise.  Note as a corollary that in systems with both electrons and holes, as in excitonic insulators and compensated semimetals, one can have macroscopic phase separation as in electron-hole droplets, but in a steady state.  This has been found experimentally\cite{RTEHD}.

But not only gas-liquid transitions are involved.  Peaks in the density of states (DOS), Fermi surface nesting, filled Landau levels\cite{RSMCondon,RSMCondon2}, and Mott transitions all tend to be optimized at particular dopings, leading to a stable ordered phase.  If the doping is tuned away from one of these special values, there is a range of energies where that same phase persists, but the gap gradually closes.  If there is a second special doping, one can have a first order transition between them, which in an electronic system becomes a NPS.  In cuprates, we find the stable dopings are $x=0$ for the AFM insulator and the VHS doping\cite{VHSnps}.  There are now many articles and conference proceedings devoted to the topic of NPS.\cite{VHSrev,phassep1,phassep2}

\subsubsection{Stripe phases}

The best-known example of NPS in cuprates is the stripe phases\cite{VHSrev,Kivelsonstripes}.  Shortly after the experimental discovery of the stripe phase in cuprates\cite{Tranq1}, we developed a NPS model of the stripes\cite{RMsuperVI,RSMstr}, postulating an electronic free energy cubic in doping, with a cusp-like minimum at half filling $x=0$ (characteristic of a fully gapped phase) and a parabolic minimum at a doping assumed to be $x_{VHS}$, where the saddle-point Van Hove singularity (VHS) crosses the Fermi level.  At the time $x_{VHS}$ was estimated to be $x_{VHS}=0.25$, but is is now known to vary somewhat depending on the cuprate family.  This results in a pattern of stripes with AFM domains separated by 2$a$-wide charge stripes (also known as bond-centered stripes).  Here $a$ is the in-plane lattice constant of the nonmagnetic parent phase. The reason for the two-atom width is that a remnant of the $(\pi,\pi)$ VHS cannot be confined on a 1$a$-wide stripe.

The model is in good agreement with experiment, and its predicted minigaps in the dispersion perpendicular to the stripes are consistent with recent density functional theory (DFT) results on a related stripe material, the superconducting nickelates\cite{Nick}.  Since all doped holes are assumed to go on the charged stripes which repel each other, doping consists of adding more charged stripes, which always stay as far apart as possible.  However, numerical calculations showed that the holes spread off of the charge stripes, so the actual charge modulations are considerably less than the ideal $x=0$ on AFM stripes to $x=0.25$ on charge stripes.  [Note that this is in direct analogy to the spread of the magnetic fields off of the vortex cores in a type II superconductor, so that the actual field modulation is never larger than the lower critical field $H_{c1}<<H_{c2}$, the highest critical field.]  

The phase near 1/8th doping is special: since the AFM stripes also need to be at least 2$a$ wide, the narrowest possible stripes are $P_c=4a$ wide -- technically, charge periodicity $P_c=4a$ but magnetic periodicity $P_m=8a$.  So what happens when doping is further increased?  Two models were proposed:\cite{RSMstr} the simplest option is that the AFM stripes remain fixed at $2a$ in width while the charge stripes get wider, but this had not been observed experimentally, so a double NPS was proposed, where the $P_c=4a$ stripe becomes stable over a finite doping range, coexisting with another phase at higher doping.  Below we shall see that features of both models are observed experimentally.

\subsubsection{Points of accumulation}

The study of NPS provides severe challenges to theory.  One must first develop an accurate model for the uniform phases of the system as a function of doping -- knowing full well that most of these will not agree with experiment.  Next, one must determine the low-energy phases between which NPS occurs.  For conventional macroscopic phase separation, the problem would be finished, as the individual domains are so large that only the two end phases are present. For NPS one must instead determine the topology of the minority phase and how it evolves with doping.  Fortunately, DFT calculations have now developed to the point that they can detect these textured phases and provide valuable information on their shape.

In complex analysis, a holomorphic function is a function that is analytical everywhere except at a finite number of poles.  An  accumulation point is, loosely, a point that has an infinite number of poles in its neighborhood.  Such a point forms a natural boundary for a holomorphic function, beyond which it cannot be analytically continued. 

If we think of the allowed phases of a material as the poles of some generalized free energy or susceptibility, then we have found that both cuprates\cite{YUBOI} and nickelates\cite{Nick} contain accumulation points in the vicinity of an AFM order.  In YBCO$_7$, DFT finds more than 20 competing low-energy phases, mainly stripe-like, with a minimum energy phase having a charge periodicity $P_c$ of 4 copper atoms, $P_c=4a$.  Note that in a DFT calculation, each texture, or stripe configuration, counts as a separate phase, requiring a new DFT calculation with the correct unit cell.  While we only calculate a finite number of phases, these are stripe-like phases, and it is clear that as the stripe periodicity goes to infinity the phase must converge to the AFM phase.  Notably, the accumulation point arises for phases with the largest magnetic moment, so even in overdoped YBCO$_7$, the driving force for stripe formation remains magnetic -- i.e., these are all textures of an underlying AFM.

What does it mean for a material to have an accumulation point, and why does that preclude analytic continuation?  The most common reason for breakdown of analytic continuation is a first-order transition between free-energy minima.  Knowing the local structure of the free energy near one minimum tells us nothing about the second minimum.  Thus, the textured phases are a particular form of NPS, and the DFT result is consistent with our earlier picture of NPS, with the added information that the texture ends when the underlying order terminates -- as assumed in paper 1\cite{Paper1}.  To complete the circle: if the stripe phase is the nanoscale version of a first order transition, then the confined second phase could be anything, thereby giving a physical explanation why an accumulation point is a natural boundary for analytical continuation.

\subsubsection{Analogy: type II superconductor}

We assume that readers are familiar with the notion of type I and type II superconductors in a magnetic field (see, e.g., Tinkham\cite{Tink}), and show how type II superconductors constitute a representative example of topological defect induced NPS.  It is convenient to start with a type I superconductor, and introduce a (tunable) analog to long-range Coulomb interaction in a doped AFM.

First, in an infinite system, superconductivity and a normal phase in a magnetic field are two competing phases that are immiscible (Meissner effect), and so are separated by a first order transition.  For a finite sample the situation is more complex.  For definiteness we consider an oblate spheriodal type-I superconductor in a field oriented along the short axis.  If the field is excluded from the sample, it is also excluded from a shadow region both in front of and behind the ellipsoid, as the field lines spread out and come together.  That is, the energy cost of the excluded magnetic field scales as $L^3$, where $L$ is the transverse diameter of the ellipsoid, while the superconducting energy gain scales as $L^2Z$, where Z is the thickness of the ellipsoid.  Thus as $Z$ decreases the superconductor becomes less stable.  

Rather than disappearing in a first order transition, superconductivity can persist in an intermediate phase\cite{Huebener}, where the sample breaks up into finite domains of superconducting or magnetic order.  As $Z$ continues to decrease, the magnetic domains continue to shrink in diameter and crowd closer together, until the sample transforms to the mixed phase of a type II superconductor with quantized magnetic vortices\cite{Tinkham} -- a texture based on field-dressed topological defects (vortices) of the superconductor.  Since there are an infinite variety of textures, this involves an accumulation point of the free energy.  

The loss of analytic continuation arises since the normal phase can be any phase that supports a magnetic field.  Thus in cuprates, there is evidence that the superconducting vortex-confined `normal phase' is actually the stripe phase, which is enhanced in an external field, with larger correlation length.  Indeed, new phases can form due to the confining cylindrical vortex -- STM studies find two-dimensional stripes not seen in zero field\cite{JH}, consistent with DFT calculations\cite{YUBOI} and with the electron-like Fermi surface seen in quantum oscillation experiments\cite{Seba}.    

Thus, we are proposing that the textured phases are a particular form of NPS, that encompasses both cuprates and superconductors. Just as in superconductors, cuprates have a crossover as a function of dopant mobility, from type I texture, essentially macroscopic phase separation, for high mobility, to type II texture, NPS, when dopants are frozen in the lattice, where holes from a second phase are trapped on topological defects of the primary AFM phase.  We note that Krumhansl and Schrieffer\cite{KrumS} first proposed the idea that topological defects, despite their high formation energy, could still play a role in the thermodynamics of a phase.

\section{The last piece of the puzzle: XY model of cuprate CDWs}

\subsection{\bf Model of CDW-like phase}

\begin{figure}
\leavevmode
\rotatebox{0}{\scalebox{0.40}{\includegraphics{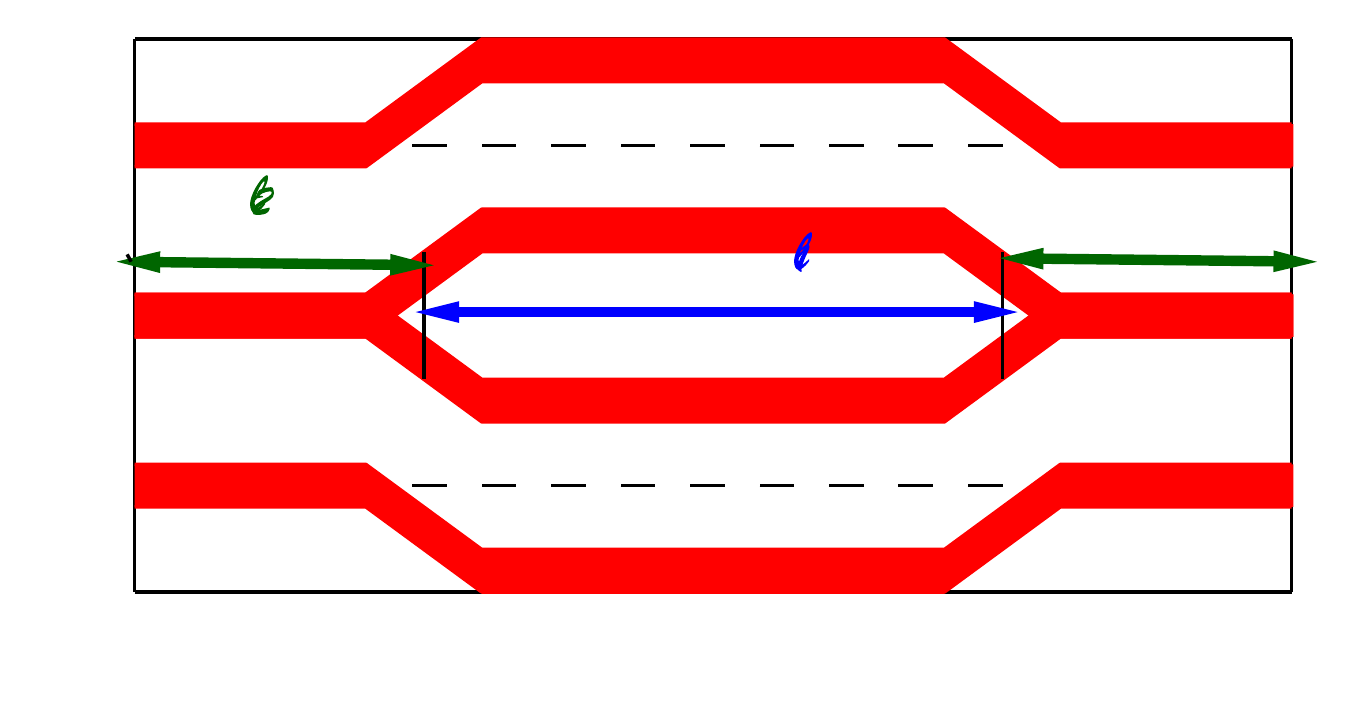}}}
\rotatebox{0}{\scalebox{0.20}{\includegraphics{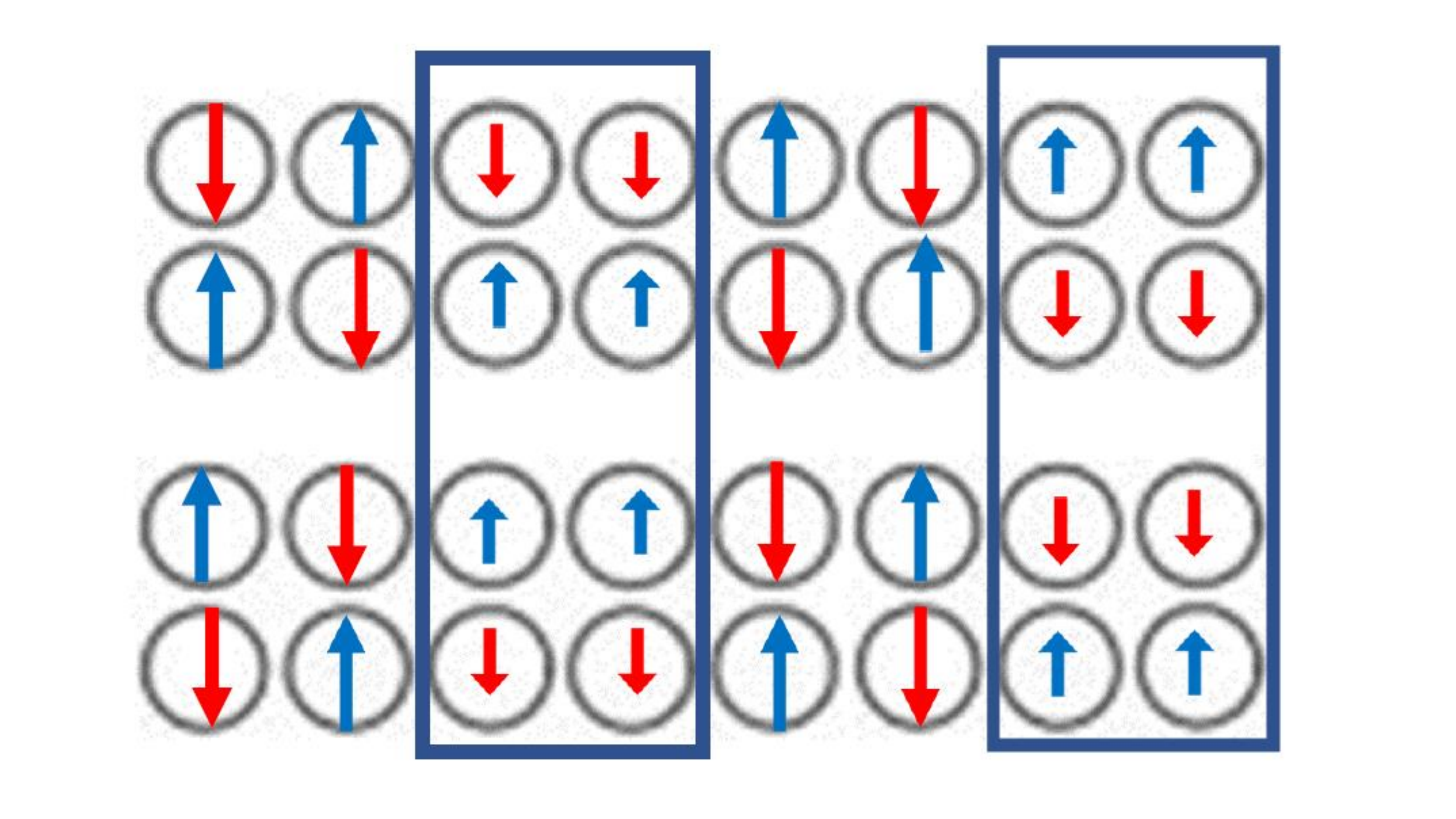}}}
\vskip0.5cm
\caption{
{\bf (a) Schematic model of CDW-like domain wall}.  (b)  Two domains of short-range magnetic order in $P_c=4$ stripes with charge stripe order.  Charge stripes denoted by boxes.
}
\label{fig:7}
\end{figure}
Here we sketch a simple model of how the CDW-like phase can evolve from the $P_c=4$ stripe phase. Figure~\ref{fig:7} shows a schematic of a segment of $P_c=4$ stripe with an extra bubble growing on one charge stripe, leading to a mix of $P_c=4$ and $P_c=6$ stripes.  [Note that odd-$P_c$ stripes tend to have higher energy\cite{YUBOI}, so we assume mixing just involves even $P_c$ values.]  If this is repeated throughout the lattice, with an average length of $l_1$ for the $P_c=6$ region and $l_2$ for the $P_c=4$ region, then the average width of a unit cell is $\bar P = (6l_1+4l_2)/(l_1+l_2)$, so that $Q=2\pi/\bar Pa$ can match the observed\cite{EHud2} incommensurate vector over a range of doping.   That is, the incommensurate stripe phase can be accomplished by generating vortices along the domain walls -- i.e., {\it by inducing an Ising-XY transition}.  This would explain the abrupt disappearance of stripes before the pseudogap collapse, since in the transition the topological defects change from domain walls to point vortices.  

In principle, since the charge stripes and spin stripes have the same width in the $P_c=4$ phase, we can interpret Fig.~\ref{fig:7}(a) as the growth of extra charge stripes for doping $x>1/8$ or (treating the spin stripes as red), the growth of extra spin stripes for $x<1/8$.  
However, this works in the Slater regime since the energy gained by nesting is larger than the energy lost in creating the vortex-antivortex pair at the ends of the bubble.  It is unlikely to work in the Mott regime for $x<1/8$, and we suggest a more likely scenario is the loss of full rows of charge stripes -- a one-dimensional analog of intercalation in graphite.

\subsection{Experimental evidence for Ising-XY transition}
In a detailed STM analysis of the incommensurate stripe phase in Bi2212, Mesaros {\it et al.}\cite{Nematic} found that the main topological defect is the merging of two charge stripes into a single one, as suggested in Fig.~\ref{fig:7}. However, in analyzing the result we find an additional subtlety.  Here we discuss this analysis.

We first summarize the STM results in more detail.  STM studies find a commensurate-incommensurate transition at $x_{CI}\sim 0.14$, vertical black line in Fig.~\ref{fig:6a}(a), from commensurate $P_c=4$ stripes to incommensurate stripes\cite{EHud2}.  The Fourier transforms of the latter consist of three dominant $q$-vectors, $Q$ associated with Fermi surface nesting, $Q_{BZ}=2\pi/a$ associated with the nonmagnetic Brillouin zone, and $S=Q_{BZ}-Q$.\cite{EHud1}  Ref.~\onlinecite{Nematic} focused on $S$, treating the $x$ and $y$ components separately.  We will focus on $Q_x$, assuming that the same underlying physics applies to all components.  In this case, $Q_xa/2\pi = 0.24$, close to the ideal 1/4 for the $P_c=4$ stripes.  Thus, we can treat the system as a nearly ideal $P_c=4$ phase, where the topological defects are the domain walls {\it of the $P_c=4$ phase}.

For an electronic phase growing on a single crystal, all of the $x$ directed stripes will have the same Bravais lattice, with only the arrangement of atoms within a cell differing on different grains [i.e., which atom is in the top left corner].  Thus there can be 16 inequivalent domains for $P_c=4$ stripes, one for each copper atom in the $8\times 2=16$ copper unit cell.  For any direction of interface, this should lead to $15\times 16/2=120$ distinct domain walls, yet the experimental situation appears much simpler.  We propose that the stripes are in a regime where the charge stripes are (quasi) long-range ordered, and the domains are associated only with spin disorder.\cite{Tranq}  In this case there are only two inequivalent domains, as in Fig.~\ref{fig:7}(b), and the problem maps onto a simple AFM on a bipartite lattice, by treating the two adjacent Cu atoms on a (bond-centered) charge stripe as a single spin. The up spins are on the A-sites in domain 1  and on the B-sites in domain 2.  Replacing the spins on all B-sites by effective spins rotated 180$^o$ in the CuO$_2$ plane replaces the AFM problem by an equivalent FM problem, which is easier to describe.  We can thus use some of our (unpublished) results on domain walls in La$_2$CuO$_4$.  

\begin{figure}
\leavevmode
\rotatebox{0}{\scalebox{0.50}{\includegraphics{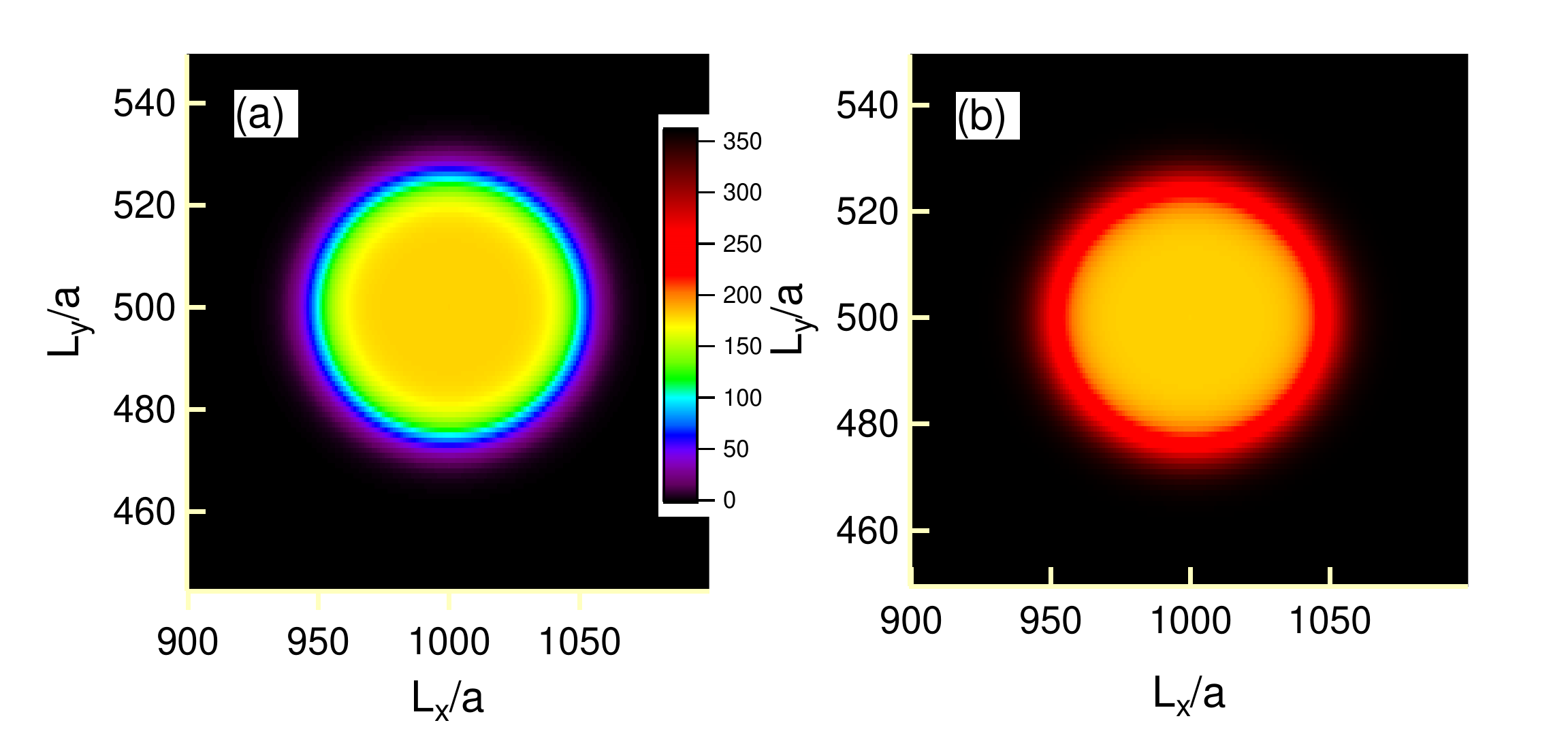}}}
\rotatebox{0}{\scalebox{0.55}{\includegraphics{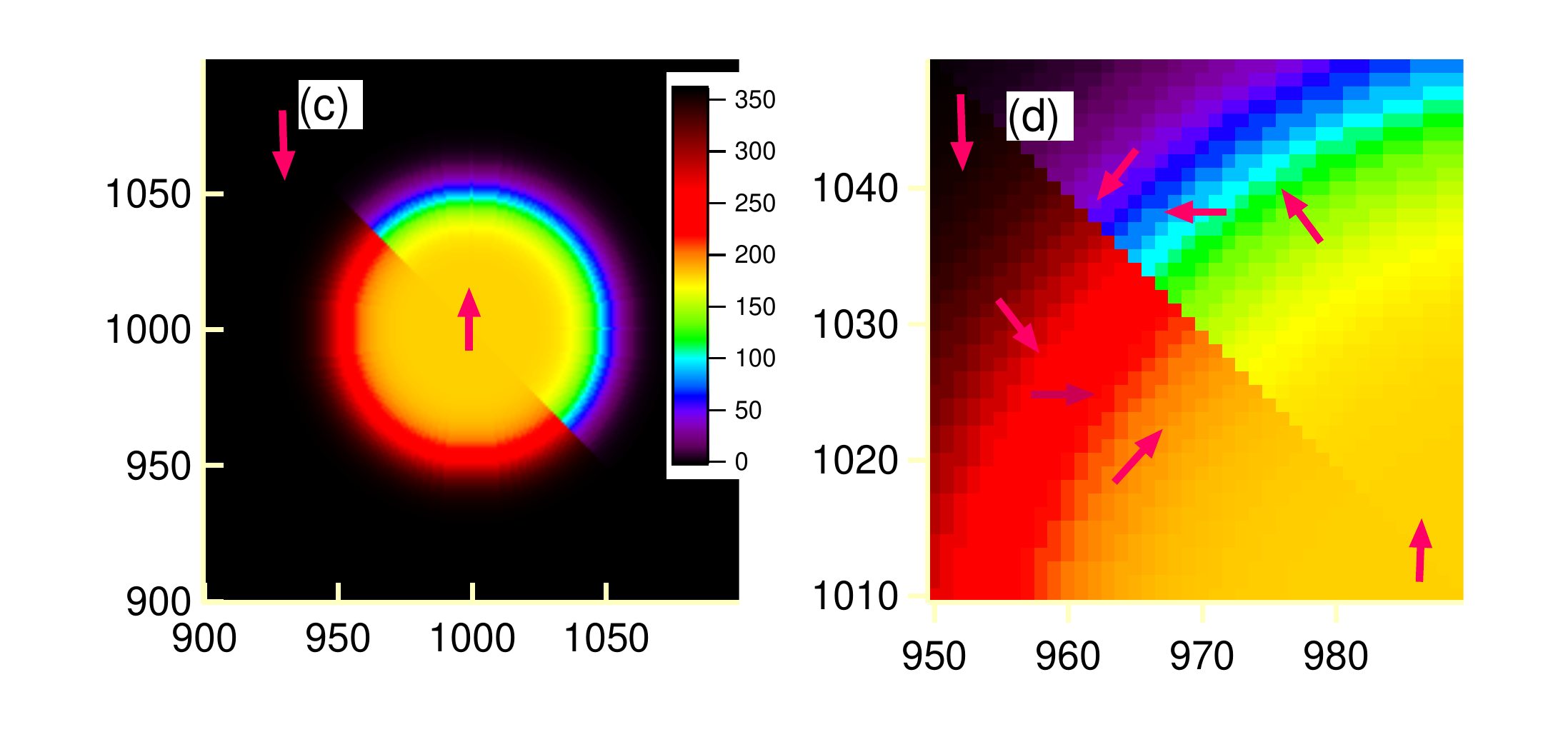}}}
\vskip0.5cm
\caption{
{\bf Models of AFM domain walls}.  (a,b) Chiral pair of uniform domain walls. (c) Domain wall with Bloch line. (d) Blow-up of (c) illustrating a vortex along the Bloch line.
}
\label{fig:8}
\end{figure}

Ref.~\onlinecite{Nematic} fit the CDW to the form $cos(S_xx+\phi(x,y))$ and then plotted the resulting map of $\phi(x,y)$.   We find that $\phi$ can be interpreted as the local spin angle.  This is illustrated in Fig.~\ref{fig:8}.  Frames (a) and (b) illustrate the simplest situation of an island of B-domain in a sea of A-domain, to be compared to the phase field maps in Fig. 2C of Ref.~\onlinecite{Nematic}.  The spin order corresponds to $\phi=0$ in A and $\phi=180^o$ in B.  The lowest energy walls are a chiral pair of Neel walls where the spins rotate by $\pm 180^o$ in plane when crossing the wall (SM Section SM-A.1).  Figure~\ref{fig:8}(a) bears a close resemblance to one experimental domain in Fig. 2C of Ref.~\onlinecite{Nematic}, which lacks any vortex pairs along the wall, except that the experimental wall is not circular, due presumably to local pinning. The more common case in Fig. 2C of Ref.~\onlinecite{Nematic} is that the walls contain a string of several similar vortex-antivortex pairs.  Figure~\ref{fig:8}(c) shows a more complicated situation in the theory, where half of the wall rotates clockwise,  the other half counterclockwise.  We note the appearance of vortices at the two crossover points.  One difference between the domain walls in the $P_c=4$ stripes and those in an undoped cuprate is that the vortex-antivortex pairs in the former are associated with defects in the charge stripes, Fig.~\ref{fig:7}(a), whereas in the latter they are associated with a Bloch line in a 360$^o$ ring wall\cite{MagDom}.

Thus, the results of Ref.~\onlinecite{Nematic} are consistent with our CDW model, Fig.~\ref{fig:7}(a), while the fact that all incommensurate data in the high doping range show the same evolution of $Q$, $S$ values leaves little room for a  distinct CDW phase.  

The data of Ref.~\onlinecite{Nematic} display a distinct excess of domain walls of one chirality.  
If this proves to be a general result, it could explain the Kerr effect\cite{Kerr}, which onsets at a temperature close to the temperature $T_H$ where stripes form, and give insight into the thermal Hall effect, which onsets below $x^*$\cite{Taill1}.  We note that in a different AFM with Neel walls a number of related anomalies are found, such as anomalous Hall effect, planar Hall effect (electric field parallel to magnetic field), and magnetically-induced chirality.\cite{chirAF}

Domain walls play a number of distinct roles.  While this paper focuses on their role as a template for confining competing phases in nanoscale textures, the Mesaros {\it et al.}\cite{Nematic} results are more concerned with the nonequilibrium state a system is left in after a quench across a phase transition, where many metastable domains are formed and the resulting topological defects have not had time to exit the system.  Thus, there have been a number of studies of the Ising-XY transition\cite{IsXY} and related models such as the fully-frustrated XY model\cite{ffXY}, but their possible role in forming textures is largely unexplored.  We note that Fig.~\ref{fig:7} offers a distinct illustration of confined vortex-antivortex pairs, which is also evident in Ref.~\onlinecite{Nematic}.

The present results raise a number of questions for future cuprate studies.  First, note that once vortices have been formed, the nesting $Q$ can be tuned just by varying the $l_1:l_2$ ratio, without inducing additional vortices, thereby suggesting a rapid C-I crossover.  On the other hand, whenever $Q=1/2n$, with integer $n$, the resulting stripes can form a commensurate domain wall array without vortices, suggesting a possibly rich doping dependence.  Unfortunately, the AFM correlation length decreases both by going into the Slater phase\cite{MBMB} and by increasing doping\cite{Ouazi,Alloul,Ghir} or $|t'|$, so this may be hard to see.

In the following section, we will summarize various experimental and theoretical data that we believe support the present picture.
Figure~\ref{fig:6a} provides a convenient summary of the experimental data we will discuss.

\section{Comparison with experiment}

\subsection{$T_H$ and the loss of domain walls}

Hall effect studies of YBCO\cite{Lifshitz} found a number of pseudogap-related phase boundaries, labeled $T^*$, $T_H$, $T_{max}$, and $T_0$, Figure~\ref{fig:6a}, which are believed to be representative of most cuprates (the termination of $T_H$ near $x=0.16$ is further discussed in Ref.~\onlinecite{CDWsend}). Here we focus on the first two.  If $T^*$ is the onset of short-range AFM order, as we propose, then $T_H$ has a natural interpretation as the Ising-$X-Y$ transition, where domain walls are the allowed topological defect only below $T=T_H$.   As the AFM phase approaches the critical doping from below, the correlation length divergence signals the broadening of the domain walls.  When the walls get broad enough, the system will transform from the Ising to an X-Y or Heisenberg model, and the walls will be destabilized.  In this range any signal of a stripe or CDW phase is lost.   This is consistent with a recent proposal that critical fluctuations of the pseudogap melt the charge order.\cite{COmelting}  The Ising-XY transition was directly observed in early experiments on cuprates\cite{Birg}.

\subsection{Similarity of stripes and CDWs}

If stripes and CDWs have independent origins, one would expect them to have different properties -- CDWs without AFM stripes, different harmonic content, etc.  But that is not the case.  First principles calculations\cite{YUBOI} shed light on this, finding uncharged antiphase domain walls in YBCO$_6$ that evolve into charged domain walls in YBCO$_7$.  The uncharged walls are stripe-like, being purely repulsive and staying as far apart as possible, whereas the charged walls are CDW-like, having a minimum energy for a finite charge periodicity $P_c=4$ (corresponding to 4 Cu atoms) which is approximately consistent with Fermi surface nesting.  However, these CDWs are highly unconventional, since the dominant contribution to minimizing their energy involves maximizing the average magnetic moment. Thus, we propose that the stripe-CDW crossover consists of rearranging AFM domain walls, from staying as far apart as possible to adjusting their average spacing to match the Fermi surface nesting.

Similarly, STM studies look almost indistinguishable through the full doping range, and a clear stripe-CDW transition is not often found\cite{EHud1,EHud2,patches,JennyO}.   While STM experiments are not sensitive to magnetic order, RIXS experiments find spin wave spectra (paramagnons) over the full pseudogap doping range\cite{Ghir,LeTacon}.  Finally, the original quantum oscillation (QO) evidence for a CDW phase is now believed to be caused by the $P_c=4$ stripe phase.\cite{Ramshaw}  These results are discussed further below.

\subsection{$P_c=4$ stripes and quantum oscillations}
\begin{figure}
\leavevmode
\rotatebox{0}{\scalebox{1.0}{\includegraphics{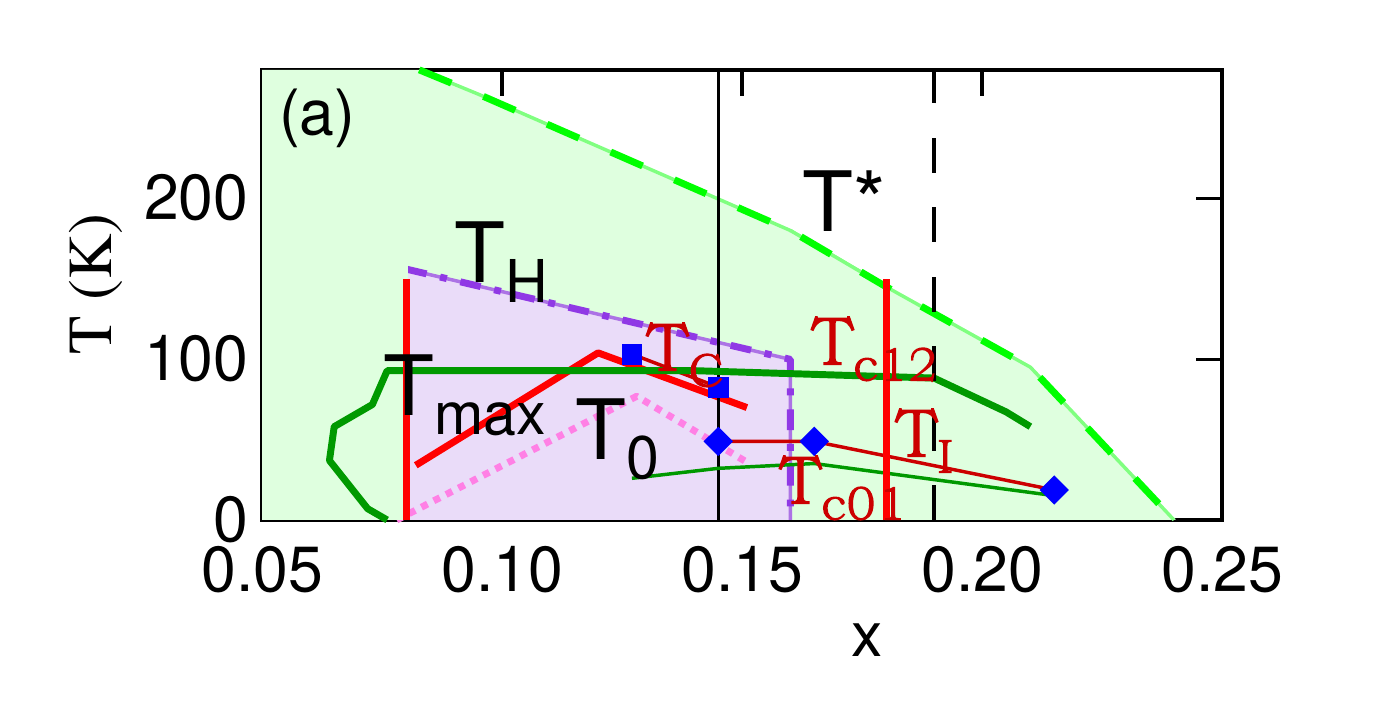}}}
\rotatebox{0}{\scalebox{0.50}{\includegraphics{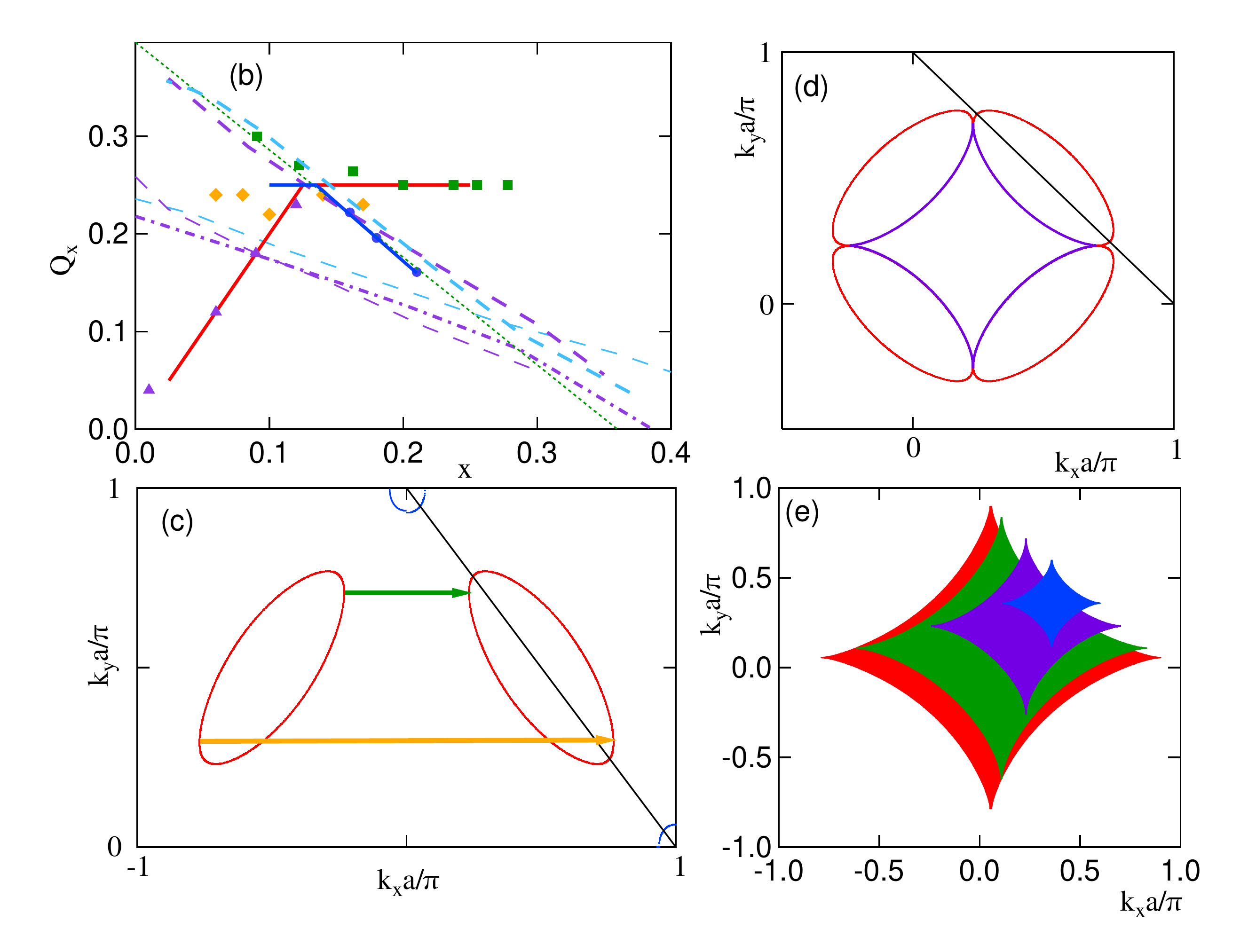}}}
\vskip0.5cm
\caption{
{\bf Domain wall evolution}.  (a) Pseudogap phase diagram of YBCO\cite{Lifshitz}, compared to STM results on Bi2201\cite{EHud2}.  For YBCO, $T^*$ is the pseudogap onset, which we interpret as short-range $(\pi,\pi)$ AFM order, while we interpret $T_H$ as the Ising--X-Y transition, where domain walls become unstable, while $T_{max}$ and $T_0$ refer to the appearance of patches of charge period 4 1D (or 2D) stripes.  For Bi2201 the blue squares ($T_C$) and diamonds ($T_I$) track the peaks in the patch maps of commensurate and incommensurate stripe phases in different samples.  The green curves labelled $T_{c(n-1)n}$ are the superconducting $T_c$s of Bi22(n-1)n, $n=1,2$.  The red vertical lines indicate the dopings at which the QO effective mass diverges\cite{Ramshaw}, presumably associated with percolation of the $P_c=4$ stripe phase.
(b) Charge order wave vectors $Q_x$, in units of $2\pi/a$, for Bi2201 (purple triangles\cite{TranqBi} and blue circles\cite{EHud2}) and Bi2212 (orange diamonds\cite{per4} and green squares\cite{GhirQ}).  Note that the orange diamonds represent the average of $Q_x$ and $Q_y$, although the differences are small.  Green dotted line is a guide for the eye for the $S_2$ branch.  Thick light-blue lines represent models of Fermi surface nesting for NM phase (dot-dashed line) or AFM hot-spot nesting (dashed lines). Thin blue and violet lines represent an alternative AFM nesting, for two model dispersions -- see SM Section SM-C.
(c-d)  {\bf Hot-spot nesting for AFM with gap $\Delta=$150~meV.}  (c) Nesting vectors $Q$ (green arrow) and $S$ (orange arrow).  (d) Double nesting, leading to star-shaped Fermi surface (violet).  (e)  Evolution of Fermi surface area with doping, $\Delta$ = 250 (blue), 150 (violet), 50 (green), and 20~meV (red).
}
\label{fig:6a}
\end{figure}
\subsubsection{QOs}
In all of the above discussion, the $P_c=4$ stripe phase has played a special role.  At low doping, it is the narrowest possible stripe phase, while at high doping it forms the lower branch of the commensurate-incommensurate (C-I) transition.  In Section III.B we saw the important role played by the $P_c=4$ stripes in the CDW formation.  Here we compare data from several different experiments to demonstrate that the doping evolution of this phase is most consistent with a percolation picture associated with NPS.    

The original evidence for CDWs in cuprates came from QO measurements of a new Fermi surface, which was found to be a small electron-doped pocket, much smaller than the expected AFM pockets, with an area corresponding to 0.04 holes per Cu.\cite{Seba}  However, this single Fermi surface is found to persist over a doping range $0.08<x<0.18$ (vertical red lines in Fig.~\ref{fig:6a}(a)), while changing its area by only 20\%.\cite{Ramshaw}  Such behavior is not consistent with a homogeneous electronic system.  The observed area change can account for only 1/10 of the doping change, and if there is a second component that accounts for 90\% of the doping, it is highly unlikely that the small Fermi surface could avoid much larger changes.  Thus the simplest interpretation is that the $P_c=4$ phase is involved in NPS, specifically that it is uniform at precisely 1/8 doping, but away from that doping it develops topological defects (both for $x<0.125$ and $x>0.125$) that increase in number with doping.  For qualitative purposes, one can introduce a simplified percolation model with only three phases, $x_0=0$ (corresponding to doping on the magnetic stripes), $x_1=0.25$ (on the charged stripes) and $x_{QO}=(x_0+x_1)/2 =0.125$ for our stripe phase.  Then 2D percolation of the stripe phase would occur when the stripe phase population falls to 50\% of the total, or at $(x_0+x_{QO})/2=0.0625$ and $(x_1+x_{QO})/2=0.1875$, close to the experimental range\cite{Ramshaw}.  Indeed, experimental evidence for anomalies at these dopings has been proposed.\cite{Ando2}  Moreover, hopping on the percolation backbone near the limits of percolation could explain the observed effective mass divergence.\cite{Ramshaw}  Such a NPS of the $P_c=4$ stripes had been predicted earlier\cite{RSMstr}, and we shall see below that it seems to be the best way to explain apparently contradictory experimental results.  Similar QOs have been found in the Hg cuprates, but with a larger area.\cite{GrevenQO}

\subsubsection{STM}
STM studies by Mesaros {\it et al.}\cite{per4} confirm this picture, finding the clear presence of patches of the $P_c=4$ phase over the full doping range.  While they ascribe this to discommensurations in a CDW phase, they don't address the issue of why this phase persists without change over such a wide doping range.  We believe that the NPS model affords a better description of the relevant physics, including percolation effects. 

However, if this is NPS, then experiments should see some clear evidence for the second phase or the corresponding topolgical defects.  Other groups have taken advantage of STM's ability to effectively erase one source of heterogeneity, to study a less disordered sample.  It is typically found that sample surfaces consist of arrays of gap maps -- roughly 3 nm patches with distinctly different combined pseudogap-superconducting gaps covering the surface.\cite{patches}  Areas with a common gap size can be masked off and are found to correlate with a distinct  local doping\cite{EHud1}, associated with clusters of a particular interstitial oxygen impurity\cite{JennyO}.  The distinct gap sizes are associated with distinct $q$-vectors, proposed to match a nesting vector of the underlying Fermi surface\cite{EHud1} -- that is, the $q$-vector is incommensurate and changes with doping.  How does NPS reconcile these results?  Within the $P_c=4$ percolation limits, most patches have this periodicity, while the remaining patches are spread over many other periodicities.  By sorting these remaining patches, one can detect a large variety of alternate, incommensurate periodicities.

In an extension of this work\cite{EHud2}, a well-defined commensurate-incommensurate transition of the stripes was found in Bi2201.  The vertical black line in Fig.~\ref{fig:6a} defines the doping $x_{CI}$, where the charge order crosses over from commensurate stripes at $P_c$=4~Cu to incommensurate $P_c>4$.  At each doping the gap size distribution was plotted, and it displays a well-defined peak.  We assume that the most common gap size correlates with the dominant domain wall order at that average doping (blue symbols in Fig.~\ref{fig:6a}  -- see SM-A.2 for details). The blue squares in Fig.~\ref{fig:6a} define the dopings at which a commensurate $P_c=4$ stripe order is observed, whereas the blue diamonds indicate incommensurate order, with average periodicity between $P_c=4$ and $P_c\sim 6$.  Consistent with this interpretation, we see that the blue squares have their highest binding energy at $x=1/8$, where the $P_c=4$ stripes should be a uniform phase.  We note further that at the C-I transition the gap size decreases discontinuously, consistent with the Mott-Slater transition\cite{MBMB}, and concurrently, exactly at the C-I transition, the gap distribution has two peaks.

\subsubsection{Hall effects}
Here, we add data from Hall effect experiments to round out the picture of this phase.  We focus on the phase boundaries, $T_{max}$ and $T_0$, Fig.~\ref{fig:6a}(a), determined in a Hall effect study of YBCO\cite{Lifshitz}.  Since these boundaries are correlated with a negative component of the Hall density, they were assumed to be a signature of the same Fermi pocket found in QO studies.  Thus, $T_{max}$, which signals when the (positive) Hall coefficient $R_H$ has a maximum, is assumed to be the point at which this new phase first appears, and $T_0$, where $R_H=0$, indicates when it begins to dominate the spectrum.  The near agreement of $T_{max}$ with the STM onset temperature $T_C$, Fig.~\ref{fig:6a}(a),  suggests the correctness of this assignment.

The high-doping end of the $P_c=4$ phase is more complicated.  First, $T_{max}$ in YBCO and $T_C$ in Bi2201 both terminate abruptly near $x=0.15$, but whereas nothing is found at higher doping in YBCO, there is an incommensurate branch $T_I$ in Bi2201.  In contrast, the QOs in YBCO extrapolate to an effective mass divergence at $x=0.18$.  One can formulate a consistent picture as follows: the $P_c=4$ phase is heading toward a percolation crossover at $x=0.18$, when this is interrupted by the C-I transition, which forms an incommensurate phase lacking the electron pocket. Note that the last doping at which the QOs are observed is $x=0.15$, consistent with this picture.

\subsection{Evolution of $Q$ vectors}
\subsubsection{Stripe phase}
The regime below $x=0.125$ was discussed in Ref.~\onlinecite{RSMstr} (see Fig.~14), in terms of the stripe evolution at low doping replaced by nanoscale phase separation (NPS) involving the $P_c=4$.   In the stripe doping range very similar neutron scattering results are found for LSCO\cite{Yamada}, YBCO\cite{TranqYB1,TranqYB2}, and Bi2201\cite{TranqBi}.

In the studies discussed above, we focused on STM and QO analyses where individual gaps were probed.  Recently there have been many studies of charge order using x-ray spectroscopies such as resonant inelastic x-ray scattering (RIXS)\cite{CoDam,GhirQ}, which can only find a volume-average nesting vector.  In Fig.~\ref{fig:6a}(b), we compare experimental data for the charge ordering vectors found in Bi2201 and Bi2212.  At first blush, the data from different experiments would seem to have little in common.  Below we first classify the results, then try to interpret them.  But first a few technical details.  We have replotted the data from Ref.~\onlinecite{Ghir}, since they derived the dopings from an outdated model of a `universal supercondicting dome'\cite{UniDome}, replacing them with values based on a dome specific to Bi2201.\cite{JennySci}  Secondly, Ref.~\onlinecite{EHud2} plotted their data as $Q$ vs gap size $\Delta$ for individual gap patches.  Rather than try to convert $\Delta$ to $x$, we sketch an approximate blue line in Fig.~\ref{fig:6a}(b), and only include blue circles for the average doping of each sample.  Lastly, neutron scattering\cite{TranqBi} observed magnetic nesting with $q$-vector $Q_{mx}=\pi-\delta_x$ and we inferred a corresponding charge vector $Q_x=2\delta_x$ (violet triangles).  These results are consistent with the corresponding data in La-cuprates (red solid lines), where stripes are more intense and the charge vector $Q$ has actually been observed, but only for $x$ = 0.12 and 0.15 -- i.e., in the $P_C=4$-phase.\cite{TranqChStr}

We note that all four experiments find doping ranges with constant $q$-vector $Q=0.25$, consistent with the $P_c=4$ stripe phase, but some experiments find additional charge order associated with the lines labeled $S_1$ and $S_2$.  $S_1$ is the well-known stripe $Q$ vector of the Tranquada model\cite{Tranq1}, consistent with charge stripes ordering by staying as far apart from each other as possible.  We focus mainly on branch $S_2$.  We note that while this branch was found in both STM\cite{EHud1,EHud2} and RIXS\cite{Ghir} experiments, and both experiments report a commensurate-incommensurate transition at the same doping $x_{C-I}\sim 0.14$ between the $P_c=4$ and $S_2$ branches, each experiment observes different segments  of the branches, and the data sets have only a single point in common, exactly at the transition.  We interpret this as follows. STM sees only data which are static on the time scale of the experiment, which can last several days, while RIXS sees fluctuating, quasi-elastic data.  Thus, the transition involves a ground state transition from $P_c=4$-stripes for $x<x_{C-I}$ to $S_2$ incommensurate stripes for $x>x_{C-I}$.  

The fact that the $P_c=4$ stripes have the same commensurate $Q$-vector over a wide doping range is consistent with the fact that the corresponding Fermi surface area seen in QO experiments scarcely changes with doping, and strongly suggests that this phase is associated with extrinsic NPS -- i.e., confined to small patches where the doping is locally close to 1/8.  

When we try to understand the physics in the low doping regime, where the $S_1$ line should be, a conundrum arises.  The patch {\it energy} seems to evolve smoothly with the average doping on the patch, suggesting that it is sensing the expected stripe order for that doping.  However, the average patch {\it diameter} is about 3~nm, or wide enough for 7.5 copper atoms. That is, the patch diameter is smaller than the stripe wavelength, which increases as doping is reduced.  The only way to reconcile these facts is to assume a `patch-Kondo' effect: the patches are sensitive to the probability of having $N_c$ = 0, 1, 2, etc., charge stripes on them, and this probability changes systematically with $x$ due to fluctuation effects.  What does this imply for the STM signal of a particular patch?  If $N_c$ = 0, there is clearly no signal, for $N_c$ = 1, the stripe can be anywhere on the patch, contributing to a uniform background.  The probability of $N_c>2$ will be small, unless the patch diameter is large compared to the stripe wavelength.  Thus, the STM signal will be controlled predominantly by the $N_c=2$ patches.  For the average patch size, two charge stripes can fit only if they are in the $P_c=4$ configuration.  This explains the STM observations: the STM signal for the average patches will always see $Q=0.25$, consistent with $P_c=4$, at low doping, while the intensity of the modulation should systematically decrease with decreasing $x$.  In principle, if one probed patches of increasing diameter, one should expect to see $Q$ systematically decrease from 0.25 towards the bulk average $Q$ vector at that doping.  There may already be a hint of this in Fig.~\ref{fig:6a}(b).  The data of Ref.~\onlinecite{per4} represent patch averages of the $P_c=4$ stripes, and they seem to fall systematically below the $Q=0.25$ line, with a slight tendency to follow the $S_1$ and $S_2$ lines.
Thus the present model can describe the $P_c=4$ and $S_1$ branches of Fig.~\ref{fig:6a}(b).

Finally, we can begin to compare the development of texture in an AFM as $x$ increases from 0 with the development in the $P_c=4$ stripe phase as $x$ increases from 0.125.  In the AFM phase of La$_{2-x}$Sr$_x$CuO$_4$, the first holes are localized on 5-Cu site clusters where they frustrate AFM coupling, leading to a loss of long range AFM order at $x=0.02$, while stripes appear around $x=0.05$.  In the $P_c=4$ stripe phase the first holes go to form $P_c=6$ bubbles, bounded by a vortex-antivortex pair, leading to the C-I transition at $x=0.14$\cite{EHud2}, the end of the $T_H$-curve\cite{Lifshitz}, possibly indicating loss of quasi-long-range stripe order, at $x=0.16$, and a percolation transition near $x=0.18$\cite{Ramshaw}.

\subsubsection{CDW-like phase and Mott-Slater transition}

We propose that the $S_2$ line is a natural extension of the $S_1$ NPS stripe model to higher doping.  If we assume that neither charge nor magnetic stripes can be less than 2 copper atoms in width, then the $P_c=4$ stripe would be the narrowest stripe order possible.  This could account for its great stability, as the charges would be modulated on the finest length scale possible, giving rise to the weakest electric fields at large distances.  In this case, the only way to add more charges to the stripes would be to invert the stripe pattern -- to fix the magnetic stripe width at two copper atoms while increasing the charge stripe width\cite{RSMstr}.  The resulting $q$-vectors would follow a straight line from $Q=0.25$ at the doping of the $P_c=4$ stripes, close to $x=1/8$, to $Q=0$ at $x_{end}$, the natural doping of the second phase on the charge stripes.  The dotted green line in Fig.~\ref{fig:6a}(b) shows a good fit of this model to the data, with $x_{end}=0.36$, close to the predicted value\cite{RSMstr} $x_{VHS}$.  [From Paper 1, $x_{VHS}$ for Bi2201 is not well known, but seems to lie in the range $x\sim 0.32-0.40$.]  This model has another interesting feature.  Recall that for the conventional stripes the $Q$ vectors satisfy $Q_m$ = $\pi(1-\delta)$ for magnetic stripes and $Q$ = $2\delta$ for charge stripes, where $\delta = x$ as $x$ varies from 0 to 0.125.  Assuming the same formula foe $S_2$, where $Q$ varies from 0.25 at $x=0.125$ to 0 at $x=x_{VHS}$, then $Q_m$ varies from 0.875$\pi$ to $\pi$ over the same range.
At the VHS, these are just the expected $Q$-vectors for intra- and inter-VHS scattering, respectively.  This provides further evidence that the pseudogap collapse is driven by the loss of AFM order.

However, this leaves open the question of why the stripes undergo a C-I transition.  It has been suggested that on the incommensurate side $Q$ is associated with a form of nesting -- either Fermi surface nesting\cite{EHud1} or `hot spot' nesting\cite{Comin}.  While nesting of the NM Fermi surface has a qualitatively similar doping dependence, it cannot quantitatively fit the $S_2$ line\cite{EHud2}, as we confirm in Fig.~\ref{fig:6a}(b), thick violet dot-dashed line.  On the other hand, it has also been suggested that the nesting might be related to a Fermi surface modified by magnetic order, and we explore this possibility in SM Section SM.C, using the AFM model of Paper 1\cite{Paper1}.  Since the Fermi surface consists of electron and hole pockets similar to those found in electron-doped cuprates, there are several possibilities, including the thin light-blue and violet curves (these refer to two different tight binding models of the same nesting, SM Figs.~S2(a) and (b)).  

We find that a form of hot-spot nesting between two adjacent hole pockets, green arrow in Fig.~\ref{fig:6a}(c), provides a good description of the full $S_2$ curve -- violet dashed line in Fig.~\ref{fig:6a}(b).  Frame (d) illustrates the result of simultaneously nesting along $x$- and $y$-axes.  If we recall from Paper 1\cite{Paper1} that the red parts of the curves are shadow bands with low intensity, the resulting Fermi surface is the star-shaped violet curve, the shape proposed for the QO Fermi surface\cite{Seba}.  In frame (d) we illustrate the evolution of the Fermi surface area as doping increases and the magnetic gap is reduced.  Note that this clears up another puzzle: why do the QOs always have the same Fermi surface area, except in Hg cuprates?  Because most represent the $P_c=4$ phase, while Hg cuprates are on the $S_2$ line.

This is a very significant result.  First, it confirms that the C-I transition is a form of Mott-Slater transition.  Secondly, it is a {\it nesting of the AFM Fermi surfaces}, proving that the stripes are excitations of an underlying AFM order.  Third, we find that there are two symmetry related nesting vectors, $Q_x$ and $S_x$, which satisfy $Q_x+S_x=2\pi/a$, consistent with experiment.\cite{EHud1}

\subsection{Other CDWs}

Many additional examples of cuprate charge order are reviewed in Ref.~\onlinecite{CoDam}.  In general, the phase diagrams are similar, in that most show a dome of charge order centered near 1/8 doping, with variations in the onset temperature and width in doping of the domes.  Despite this, the $q$-vectors are very different in different cuprates.  While most are approximately consistent with the $S_1$ and $S_2$ branches, a few are notably different.  Within the present framework, this can most plausibly be understood by noting that most of these cuprates are doped by excess oxygen, and the doping sites and oxygen mobility can vary greatly in different cuprates.  Enhanced oxygen mobility can lead to larger spatial scales of phase separation, which in turn can lead to oxygen ordering with periodicity unrelated to Fermi surface nesting. This is further discussed in the following subsection.

In addition to the incommensurate stripe phases noted above, cuprates have a number of other structural phase transitions and/or CDWs that are very material specific, tend to turn on at higher temperatures, and seem to have little to do with the pseudogap phase, until one looks more closely.

We recall that in La$_2$CuO$_{4+\delta}$ there is macroscopic phase separation, due to oxygen being mobile enough to follow the holes.  It has further been found that oxygen doping resembles intercalation in graphite: the excess oxygen intercalates between layers, and doping leads to different staging of the layers\cite{BirgO1}.  For example, stage $N$ intercalated graphite means that there are $N$ layers of graphite between successive layers of the intercalant layer.  Similarly, in La$_2$CuO$_{4+\delta}$ there are $N$ Cu layers between successive layers of interstitial oxygen.  To make the pictures of phase separation and intercalation mutually consistent, Daumas and H\'erold\cite{Daumas} showed that there are equal amounts of intercalant in each possible intercalant layer, in each layer the intercalants are phase separated into islands, and for each island there is a similar island separated by $N$ layers of the host -- graphite or copper.  Since most of the holes in cuprates will be attracted to the copper layers adjacent to the interstitial oxygens, intercalation can be thought of as a three-dimensional analog to stripes in two dimensions.  For present purposes, the most important result is what the oxygen rich islands can tell us about the `second phase' in the cuprate phase separation: these islands of optimal superconductivity are characterized by in-plane ordering of the intercalated oxygens.\cite{BirgO2}

Similar oxygen ordering phenomena are found in other cuprates.  Thus, YBCO is doped by adding oxygens to a copper chain layer.  Doping progresses mainly by adding filled chains, which then order similarly to a two-dimensional form of intercalation\cite{Fontaine}.  Along these chains, STM and ARPES studies find a form of CDW order related to hole doping along the chains\cite{chainCDW}.  Similarly, in Hg cuprates evidence has been found for chain ordering of of both oxygen dopants and Hg vacancies\cite{Hgchains}.  

To study even higher doping, several groups have turned to high-pressure oxygen synthesis of LCO-analogs with the La fully replaced by Sr or Ba.  The resulting samples have turned out to have high oxygen vacancy content, Sr$_2$CuO$_{4-\delta}$ (SCO)\cite{SCO} and Ba$_2$CuO$_{4-\delta}$ (BCO)\cite{BCO}, but with significantly increased $T_c$ values, up to 90K for SCO and 70K for BCO -- evidence for a `second dome' of superconductivity at higher doping.  Here we note only that SCO samples are multiphase, having several phases with distinct $T_c$ values, and also a number of phases with clear oxygen vacancy order.    At this point however, it is not clear whether the vacancies help or hurt superconductivity.

At first sight, oxygen or vacancy order seems to have nothing in common with our picture of a textured AFM. However, the question arises: are these phases predominantly ionic, or could the oxygen order be driven by an underlying electronic Wigner crystal?  This would then represent another form of NPS, closely related to the ferron model of a doped AFM\cite{Nagaev}, and consistent with the vortex defects of an XY model.

\subsection{More on the Mott-Slater transition}

\subsubsection{Homogeneous models}

The theory of the Mott-Slater transition needs to follow two separate paths.  First, the effect should be found in other theories of the cuprates, if they assume that the cuprate phases are essentially homogeneous.  Secondly, one must understand what traces of the Mott-Slater transition remain in a NPS model, for comparison to experiments.

When experiments\cite{TaillFS} found that the pseudogap terminates close to the doping $x$ where the VHS crosses the Fermi level, $x_{pg}\sim x_{VHS}$, several theoretical calculations\cite{CDMFT2,CDMFT1} were able to reproduce this result only for $t'>t'_{cr}$, while for $t'<t'_{cr}$, $x_{c}< x_{VHS}$, Fig.~\ref{fig:4}.  (To avoid confusion, we use the symbol $x_c$ to denote the crossover doping that was identified as the pseudogap closing in Refs.~\onlinecite{CDMFT2,CDMFT1}.)

To compare these results with experiment\cite{TaillFS}, we plot Fig.~2 of Ref.~\onlinecite{CDMFT2} in Fig.~\ref{fig:4}, making use of a symmetry of the Hubbard model (under $x\rightarrow -x$ and $t'\rightarrow -t'$) to replot the data for $t'>0$, $x>0$ (hole doped, non-cuprate like)) as $t'<0$, $x<0$ (electron doped cuprate-like). We first focus on the comparison between $x_{c}$ (solid black and green lines) and the dressed $x_{VHS}$ (long-dashed blue lines).  These are in good agreement for electron-doping ($x<0)$ into the upper Hubbard band (UHB), but for hole-doping into the lower Hubbard band (LHB) the agreement lasts only for $t'>t'_c\sim$~-0.15, while for smaller $t'$, $x_c<x_{VHS}$.  However, $x_c$ is in poor agreement with the experimental\cite{TaillFS} $x^*$ values (red circles), which are much closer to $x^*=x_{VHS}$, as discussed in Ref.~\onlinecite{Paper1}.  (We discuss in SM-A.3 how the experimental data are plotted here.)  On the other hand, $x_c$ is in reasonable agreement with $x_{C-I}$ for Bi2201 (green diamond and black solid vertical line in Fig.~\ref{fig:6a}(a)).  Finally, this result is consistent with the finding\cite{MBMB} that LSCO is in the Mott phase while other cuprates have $x_{VHS}$ in the Slater phase.

\subsubsection{Heterogeneous experiments}

\begin{figure}
\leavevmode
\epsfig{file=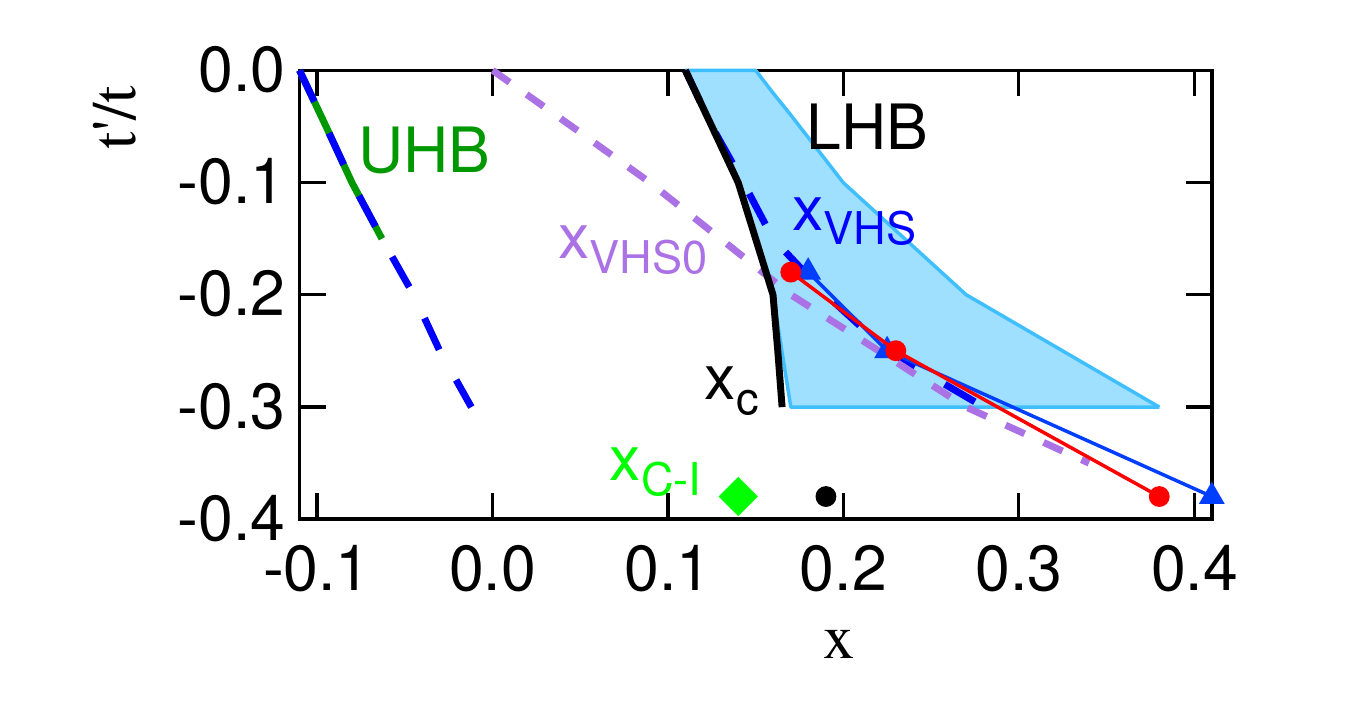,width=11.5cm, angle=0}
\vskip0.5cm  
\caption{
{\bf Comparing calculated dopings of `pseudogap' $x_c$ (solid lines) and VHS $x_{VHS}$ (dashed lines), vs $t'/t$} (assuming $t''=0)$\cite{CDMFT2}.  Blue long-dashed lines indicate $x_{VHS}$, while $x_{VHS0}$ (violet short-dashed line) is the bare VHS, which at low dopings drives the splitting of the bands into UHB and LHB; light-green diamond is $x_{C-I}$\cite{EHud2}; red dots and blue triangles represent experimental $x^*$ and $x_{VHS}$\cite{TaillFS}.
}
\label{fig:4}
\end{figure}

As we have seen above, if the cuprates remained homogeneous vs doping, they would undergo a Mott-Slater transition, leading to an emergent spin liquid phase and a large drop in correlation length across the transition,\cite{MBMB} which could be mistaken for a pseudogap collapse. However, the spin liquid is a high entropy phase, and can be avoided by phase separation.  Since NPS is present in cuprates, we ask what signatures of the Mott-Slater transition persist in the NPS regime.  Clearly, the C-I transition reflects the Mott-Slater transition, since only in the incommensurate phase is $Q$ controlled by Fermi surface nesting.  The discontinuous drop in energy at the C-I transition suggests a corresponding drop in correlation length.  We noted above the close correspondence between $x_{C-I}$ and the phase boundary $x_c$, Fig.~\ref{fig:4}.

The region between $x_{C-I}$ and $x_{VHS}$ is likely to be highly disordered, both because of frustration due to competing phases in the incommensurate regime, because of the emergent spin liquid phase near the C-I transition, and because of strong scattering associated with the VHS.  Hence the results of Ref.~\onlinecite{Tremblay} offer some confirmation of the current assignments.  There it is found that the light-blue shaded region in Fig.~\ref{fig:4} is a strange metal Fermi liquid phase, with linear-in-$T$ resistivity.  Note that $x_{C-I}$ is close to the temperature of optimal superconductivity, suggesting that AFM-CDW competition, or the emergent spin liquid phase is beneficial for superconductivity.

\section{Discussion}

\subsection{Mott physics}

An advantage of the model developed here and in Ref.~\onlinecite{Paper1} is that it is readily extendable to many other correlated materials\cite{Bian3,irSCAN,niSCAN,fSCAN,Bian1,Tranq4,Bian4}, and the small correlation length limit of this model is consistent with the SQS model\cite{Zunger}.  We thus here restate the key features.  
 
(1) We discuss electronic systems, where the doping ions are static, leading to NPS when a first order transition might be expected.  

(2) A gapped material is essentially incompressible, in the sense that uniform doping would move it away from the gap, leading to a rapid loss of binding energy.  Instead, it develops topological defects where the electrons are trapped.  The resulting texture greatly extends the range of the parent phase, much like the vortex phase in a type II superconductor.  

(3)  There are characteristic modifications of this scenario that can help confirm its correctness.  

(a) Thus in La$_2$CuO$_{4+\delta}$, the interstitial oxygen is mobile at room temperature, where the AFM order is forming.  In this case, the oxygen moves with the electrons, leading to macroscopic phase separation into undoped AFM and optimally doped superconducting domains\cite{Jorg,Bian2} -- see also Section 11.1.1 in Ref.~\onlinecite{VHSrev}.  (Recall that in LSCO, optimal doping falls close to $x_{VHS}$.)  Moreover, in the doped phase $T_c$=44K is actually higher than found in La-cuprates prepared by different doping.  

(b) In other cuprates the dopants are immobile, but have statistical clustering.  In this case, the process (3a) runs backward: the electrons pile onto the dopant clusters to form slightly more macroscopic phase separation.  That is, the extrinsic NPS patches observed in STM are a natural consequence of the intrinsic NPS.  Consistent with this picture, the patches appear to be a bulk effect.\cite{bulk}  In these materials the superconducting $T_c$ varies smoothly with doping, with clear evidence of stripes, but no sign of macroscopic phase separation.  Note that while in a uniform system the AFM phase should terminate at $x^*$, persistence of gap maps into the overdoped regime will produce a residue of AFM order at higher doping.\cite{Hussey}

(4) Cuprates display a double NPS, with one two phase regime between the $x=0$ AFM phase and the $x=0.125$ $P_c=4$ phase, and a second from $x=1/8$ to the VHS, with percolation transitions, characterized by diverging effective masses, in both regimes.  Thus the $P_c=4$ phase seen in QO measurements plays a special role in the doping evolution.

While the above features are universal, others are more characteristic of cuprates.

(5) The topological defects in cuprates are domain walls, driven by an extremely small Ising anisotropy in the exchange constant $J$.  Thus as temperatures increase, there is an Ising-to-XY transition\cite{Birg} at which the charge stripes melt, well below the pseudogap transition.  This can be identified with the curve $T_H$ in Fig.~\ref{fig:6a}(a).  Since the AFM order can be long-range only in the Ising phase, the system transforms at $T_H$ from short-range AFM order to potentially long-range CDW order (depending on the effect of disorder on the charged stripes) as $T$ is reduced.

(6) Similarly, the walls vanish at high doping, so that the final pseudogap collapse can be described in terms of the primary AFM order\cite{Paper1}.

(7) When all of these secondary manifestations are set aside, we can study the underlying AFM order.  We find that when doping $x$ and hopping parameter $t'$ are small, the AFM shows the characteristics of a Mott insulator, where a gap opens high above the temperature where long-range AFM order is found, and Fermi surface nesting plays a negligible role.  As $x$ or $|t'|$ increase, there is a non-Landau transition to a Slater phase dominated by Fermi surface nesting.  The transition is characterized by a change in the topology of the charge order and a significant increase in disorder, to the extent that the Slater phase seems to be coterminus with the strange metal phase.

(8) Correlated materials are complicated due to the presence of three forms of heterogeneity: the intrinsic NPS, an attraction of holes to dopant clustering, leading to the formation of patch maps, and the presence of an array of metastable domain walls generated by cooling the samples across the AFM phase transition too quickly (quenching).\cite{Nematic}  

The ability to map out these domain wall arrays\cite{Nematic} could prove useful for future studies of ``experimental cosmogenesis" -- specifically the Kibble-Zurek mechanism of generating cosmic strings\cite{KZ1,KZ2}.  Coming back down to earth, it would be interesting to see if the chirality of the domain walls can be tuned by cooling in a magnetic field, and if so, can this be correlated with Kerr effect or thermal Hall effect measurements?  Finally, spintronic techniques for moving AFM domain walls could be applied to eliminate one form of disorder in the pseudogap phase.

\subsection{Superconductivity}
While our focus here is on the pseudogap, we briefly note some points connected with superconductivity.  In Figure~\ref{fig:6a}(a) we plot the superconducting domes $T_c(x)$ for Bi2212 ($T_{c12}$) and Bi2201 ($T_{c01}$) as green solid curves.  For Bi2212, $T_c$ has a flat maximum extending over the doping range between the percolation limits of the $P_c=4$ phase.  In contrast, in Bi2201, $T_c$ lies close to but below $T_I$.  This suggests that in Bi2201 {\it superconductivity could be the competing phase trapped on the AFM stripes and vortices}\cite{FilSup}, an interesting complement to the high-magnetic field studies that found AFM stripes as the phase trapped on superconducting vortices.\cite{JH}  Note that for both Bi compounds the superconducting dome terminates near $x^*$.  LSCO resembles Bi2201, in that  superconducting and stripe order disappear at the same doping,\cite{LSCOstr,LSCOstr2} even when Fe-doped\cite{LSCOFe} ($x_c=0.28$).  Further, we have seen that the competing phase in oxygen-doped La$_2$CuO$_{4+\delta}$ is superconductivity.

We caution that we are here assuming a cuprate universality, as the limits of the $P_c=4$ stripe phase have been determined in YBCO, while the $T_c$ dome is from Bi2212. 

In a recent series of papers, it was demonstrated that superconducting fluctuations in most cuprates above $T_c$ cannot be explained by Ginzburg-Landau physics, but are consistent with a {\it universal} percolative superconductivity.\cite{MG1,MG2,MG3}  It was proposed\cite{MG4} that the percolation arises due to pseudogap-related inhomogeneity of the `normal' state. This is consistent with earlier proposals for optimizing superconductivity\cite{WAL,ABB}, and with the fact that optimal superconductivity falls close to the Mott-Slater transition.\cite{MBMB,hoVHS1}

\subsection{Two dimensionality}

All of the above discussion does not fully exhaust the complexity of the cuprates.  In our DFT study of YBCO$_7$, the topological defects are of two kinds: planar charged domain walls and interlayer stacking faults.  While we have concentrated on the former, a recent ARPES study of Bi2212\cite{ZXarc} finds a loss of interlayer coherence, specifically loss of the coupling between bilayer bands -- i.e., a transition to two-dimensionality. While the transition occurs in an interesting doping regime (black dot and black dashed vertical line in Fig.~\ref{fig:6a}(a) and black dot in Fig.~\ref{fig:4}), we currently have no idea whether it is connected to the C-I transition, whether it is unique to Bi2212, or how it might be related to two-dimensionality in the La-cuprates.  One hint is that three-dimensionality flattens out the VHS, cutting off its divergent peak.

\subsection{Candidates for the second phase}

Consistent with the definition of points of accumulation, we have not yet identified the second phase that is confined on the charge stripes, although there are hints that at least in some cases it involves the superconductivity.   This could explain why $T_c$ and $T_N$ often are found to be similar in this doping regime.    Since the $S_2$-branch of $q$-vectors extrapolates to charge-$Q$=0 and magnetic-$Q=(\pi,\pi)$ at $x=x_{VHS}$, another plausible candidate is the Slater-AFM.  It may be that in other systems -- Fe-based superconductors or even in the cuprates with largest $|t'|$ -- charge order plays a larger role, as discussed in SM-B.

\subsection{Comparison  with cluster-DMFT and other models} 
As noted above and in Paper 1\cite{Paper1}, our predictions differ in detail from calculations based on cluster extensions of dynamical mean-field theory (DMFT).  Here we discuss possible reasons why these differences arise.  Our DFT-based philosophy is that every material is characterized by its unique dispersion, so one needs an accurate model of the dispersion to accurately calculate the bare susceptibility.  This latter then acts as a basis for (GW-type) self-energy calculations and phase diagrams via RPA calculations, preferably with mode coupling corrections.  The DMFT philosophy is different.  It calculates a highly accurate but momentum-independent interaction corrected self energy, while cluster extensions calculate the self energy at N points, where typically $N\sim 2-8$, leading to a similarly coarse susceptibility.  Further, the dispersion has a highly simplified form.  Thus, cluster-based DMFT calculations typically use tight binding models of $t-t'-t"$ form, but with either $t'=t"=0$ or $t"=0$, neither of which are suitable for cuprates.  Specifically, for $t"=-t'/2$, the model captures the known camelsback dispersion near $(\pi,0)$, which signals that two pockets have broken off of the large Fermi surface near $(\pi,0)$ and $(0,\pi)$.  This topological transition happens at a high-order (ho) (power-law diverging) VHS, which seems to control optimal $T_c$ in the cuprates.\cite{hoVHS1}  In contrast, when $t"= 0$ the hoVHS shifts to very large hole doping, beyond the range of cuprates, and the correlation of $T_c$ and hoVHS is lost.  SM Figure S1 shows how far apart these two hoVHSs are.

 Our model also bears some similarity to a polymeric model of stripes.\cite{polymer}


\section{Conclusions}


Correlated materials combine aspects of great complexity (intertwined orders) with with other aspects hinting at great underlying simplicity -- universality, emergence of exotic phases with `colossal' properties (high-$T_c$ superconductivity, colossal magnetoresistance, heavy fermion behavior), pseudogaps, and strange metals with anomalous transport (linear-in-$T$, -$\omega$ resistivity).  For cuprates, we here identify the complexity as due to the texture that confines competing orders on charge stripes.  We have further demonstrated that the stripe-to-CDW crossover is the defining signature of the Mott-Slater transition.  Even when the physics is controlled by Fermi surface nesting, the AFM texture can adjust to accomodate nesting, while the induced vortices ultimately melt the charge stripes, leaving the pseudogap collapse to be described by a simple AFM model.\cite{Paper1}

\section*{Acknowledgements}     
This work was supported by the US Department of Energy (DOE), Office of Science, Basic Energy Sciences Grant No. DE-SC0022216 and benefited from Northeastern University’s Advanced Scientific Computation Center and the Discovery Cluster.  
We thank Adrian Feiguin for stimulating discussions.  
\section*{Author contributions}{
R.S.M. and A.B. contributed to the research reported in this study and the writing of the manuscript.
\section*{Additional information}
The authors declare no competing financial interests. 
\end{document}


\title{Mott-Slater Transition in a Textured Cuprate Antiferromagnet\\
Supplementary Material}
\author{R.S. Markiewicz and A. Bansil}
\affiliation{ Physics Department, Northeastern University, Boston MA 02115, USA}

\maketitle
\beginsupplement
\section{Domain wall phase diagram}
\subsection{Data used to construct Fig. 2}
The circular domain walls were calculated from the equation $\theta=[1\pm tanh((r-r_0)/2\xi)]\pi/2$, with $r_0=25a$ and $\xi/a=9$.  The ring is centered in a square $L=2000a$ on a side,

\subsection{Data used to construct Fig. 3}

Figure~3 includes superonducting (green lines) and STM data (blue symbols), where we use the following procedure to convert the STM gaps $\Delta$ to equivalent onset temperatures.  The superconducting gaps have been well studied, and satisfy the relation 
 \begin{equation}
2\Delta=Xk_BT_c.
\label{eq:BCS} 
\end{equation}  
For Bi$_2$Sr$_2$Ca$_0$Cu$_1$O$_6$ (Bi2201), $X$ = 12.4\cite{EHud3}, while for Bi2212 $X=10$,\cite{Vishik} with weak doping dependence.  We assume all STM gap data can be converted to temperatures by the same factor $X$.

\subsection{Data used to construct Fig. 4}
 
\begin{figure}
\rotatebox{0}{\scalebox{0.8}{\includegraphics{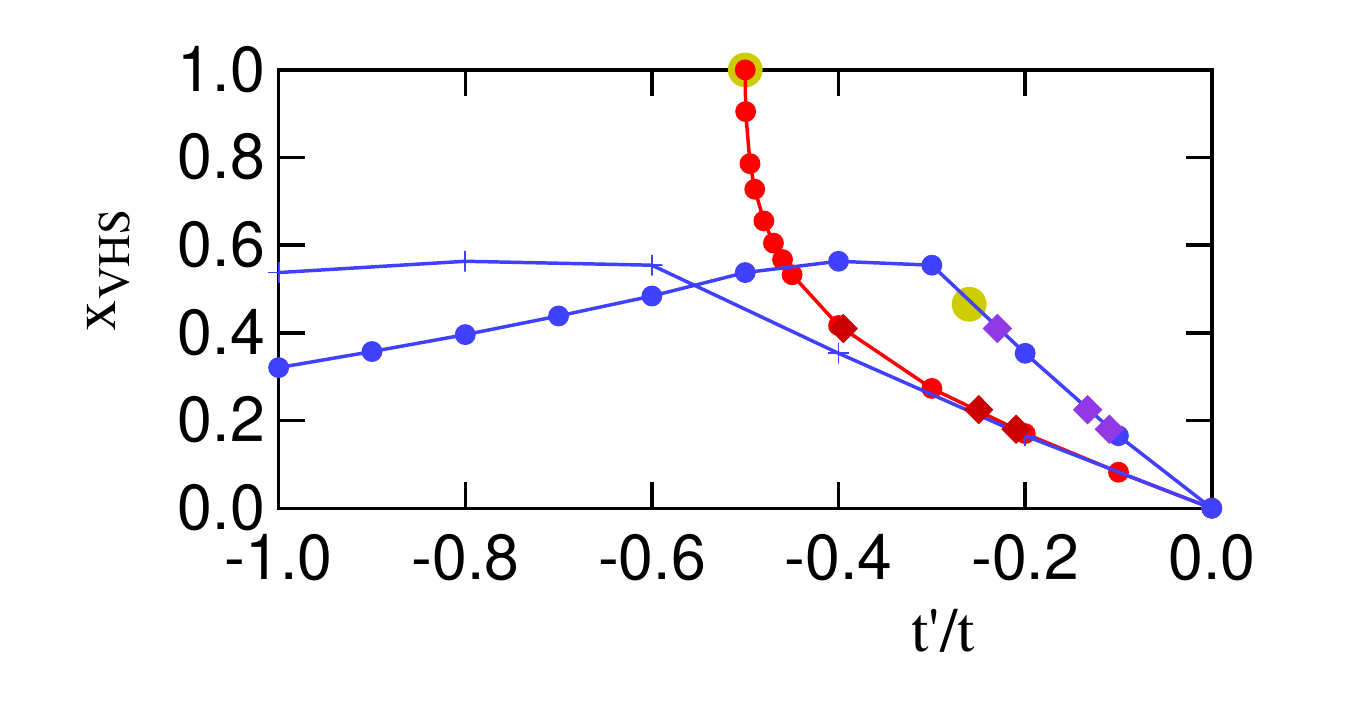}}}
\vskip0.5cm
\caption{
VHS doping $x_{VHS}$ in the $t-t'-t''$ model for $t''/t'$ = 0 (red dots) and -0.5 (blue dots).  Dark red and violet diamonds represent $t'/t$ for the data of Ref.~\onlinecite{TaillFS} within the corresponding models. Note that at low doping $t'$ for $t''=-0.5t'$ is approximately half $t'$ for $t''=0$ (blue +-signs).  Golden circles indicate hoVHSs.
}
\label{fig:A1}
\end{figure}

Ref.~\onlinecite{TaillFS} did not provide any estimate of $t'$.  To map their data onto Fig.~4, we assumed that the ARPES estimate of $x_{VHS}$ is essentially $x_{VHS0}$ and plotted the data at the appropriate $t'$.  Then Fig.~\ref{fig:A1} can be used to convert $x_{VHS0}$ to $t'$, both for the data of Refs~\onlinecite{CDMFT2,CDMFT1,Tremblay}, for which $t''=0$ (dark red diamonds), and for the parameters of the Pavarini-Andersen model, with $t''=-t'/2$ (violet diamonds). This translates into experimental\cite{TaillFS} estimates $t'_{PA}$ = -0.11 (LSCO), -0.133 (Nd-substituted LSCO), and -0.23 (Bi2201), in good agreement with other estimates.

We note one further difference between our results in Paper 1\cite{Paper1} and those of Ref.~\onlinecite{CDMFT2}, Fig.~4: in the pure Hubbard model we find the AFM VHS at the band top, so $x_{VHS}$ should extrapolate to $0^+$ as $x\rightarrow 0$.  However, a small broadening of the VHS would shift the peak to finite doping, and due to the strong divergence of the DOS, a shift to $x\sim$~0.11 would not be surprising.

\section{Hypothesis for confined phase near $x_{VHS}$}
\subsection{SO(8)}

A key question in this model is, what constitutes the second phase confined on the charge stripes?  If it arises near $x_{VHS}$, it should be one of the phases enumerated in the VHS spectrum generating algebra SO(8)\cite{SO8}.  This group is an extension of S.-C. Zhang's SO(5)\cite{SO5} model of AFM-d-wave superconductivity competition, to include additional competing phases.   In SO(8) the competing ordering phases can have ordering vectors $Q$ = $(\pi.\pi)$, $(\pi,0)$, or $\Gamma=(0,0)$.  A competing phase at $\gamma$ arises naturally in a model of inter- vs intra-VHS competition\cite{hoVHS1}. This can include superconductivity, ferromagnetism\cite{Kapit}, or CDW order.  Below we focus on CDW order, since evidence of a CDW has been found experimentally in many cuprates.
The latter is predicted to dominate for $t'/t \le -0.23$,\cite{hoVHS1} but so far there is little evidence for cuprates  with such large $|t'|$. Further, this leads us out of the strict SO(8) model, because the actual Fermi surface nesting is not at $Q=0$, when the VHS crosses the Fermi energy, but at a finite $Q$ which obeys $Q\rightarrow 0$ as $x\rightarrow x_{VHS}$.

This leads to a Hamiltonian matrix
\begin{equation}
\begin{pmatrix}
\epsilon_k & -O_{SDW} & -\Delta_d & \Pi & -O_{CDW} \\
O_{SDW} & \epsilon_{k+Q} & -\Pi & -\Delta_d & -\tau \\
\Delta_d & \Pi & -\epsilon_k & O_{SDW} & 0\\
-\Pi & \Delta_d & -O_{SDW} & -\epsilon_{k+Q} & 0\\
O_{CDW} & \tau & 0& 0& \epsilon_{k+q}\\
\end{pmatrix},
\label{eq:1}
\end{equation}
where we retain the basic notation of Ref.~\onlinecite{SO8} for ease in comparison.
Comparing Eq.~\ref{eq:1} to Eq. 1 of Ref.~\onlinecite{SO8}, we note that the elements of the spectrum-generating algebra are proportional to the various gap functions.  Thus, $\epsilon_k$ is the bare electronic dispersion, $O_{CDW}$ ($O_{SDW})$ is the CDW (SDW) order parameter, $\Delta_d$ is the $d$-wave superconductor order parameter, $\Pi$ is the pair-density wave order parameter, and we focus on the $z$-components of the vector operators $O_{SDW}$ and $\Pi$.

We note the following points.  (1) When CDW order is neglected, Eq.~\ref{eq:1} reduces to a form of Zhang's $SO(5)$, although the form of his Lie algebra looks different.  This is because the elements of the algebra can be interpreted either as order parameters or as symmetry elements, so that, e.g., $S_z$ could be interpreted as the generator of rotations about the $z$-axis, or as a component of the ferromagnetic order parameter.  We recall Zhang's prediction that if two components of the order parameter are present (e.g., $O_{SDW}$ and $\Delta_d$), then the order associated with their commutator ($\Pi$) is also present. (2) For present purposes, we are only interested in pseudogap physics, and so neglect the superconducting orders, thereby reducing Eq.~\ref{eq:1} to
\begin{equation}
\begin{pmatrix}
\epsilon_k & -O_{SDW} &  -O_{CDW} \\
O_{SDW} & \epsilon_{k+Q} &  -\tau \\
O_{CDW} & \tau & \epsilon_{k+q}\\
\end{pmatrix}.
\label{eq:2}
\end{equation}
While the original $SO(8)$ matrix is defined for the VHSs, in which case $Q=(\pi,\pi)$ and $q=0$, we extend the definition to account for the CDW nesting vector $q$, in which case $\tau$ represents a shear distortion connecting $Q$ and $q$.  (3) In this case, if $q$ is rational, $qa/2\pi=n/m$, where $m$ and $n$ are relatively prime integers, then the CDW Hamiltonian by itself becomes an $m\times m$ matrix\cite{smectic}, so self consisting in $q$ is extremely difficult.   

The picture of AFM to CDW crossover, driven by VHS competition between interVHS coupling  at $Q=(\pi,\pi)$ and intraVHS coupling at $q\sim 0$, is very appealing, and consistent with the Mott-Slater crossover.  However, it has serious flaws.  First, it does not explain the presence of the stripes.  Secondly, it does not explain why pseudogap collapse is associated with loss of AFM order\cite{Paper1}. However, it remains possible that the confined phase in the AFM texture evolves into this phase as $x\rightarrow x_{VHS}$.  Moriya\cite{Moriya} expected a similar crossover from an insulating AFM to a $q=0$ metal, but in his case the metal was a ferromagnet -- an alternate choice in $SO(8)$.
\subsection{Beyond the hoVHS}
\begin{figure}
\rotatebox{0}{\scalebox{0.4}{\includegraphics{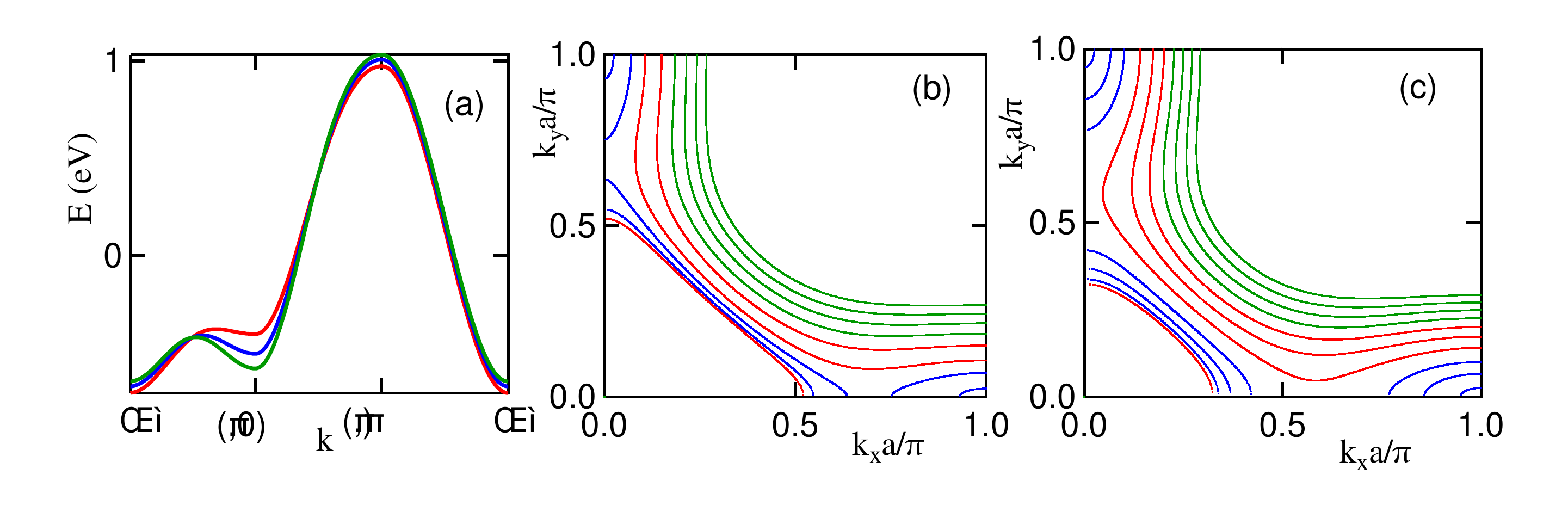}}}
\vskip0.5cm
\caption{
{\bf Doping beyond the hoVHS.} (a) Dispersions for $t'/t$ = -0.32 (red), -0.40 (blue), and -0.46 (green). (b) Fermi surfaces for $t'/t$ = -0.32 and doping (along the diagonal from top right to bottom left) $x$ = -0.03, 0.04, 0.12, 0.21, 0.31, 0.45, 0.58, 0.67, and 0.70. Blue curves are in the camelsback range where the Fermi surface has three pockets, and green curves show well-nesting regions in the antinodal regimes near $(\pi,0)$ and $(0,\pi)$.  (c)  Similar to (b), except $t'/t$ = -0.40 and doping $x$ = -0.04, 0.02, 0.10, 0.17, 0.27, 0.37, 0.52, 0.71, 0.79, 0.84, and 0.86.
}
\label{fig:A3}
\end{figure} 
Perhaps the best place to look for the competing order is in materials with larger $|t'|$, at doping beyond the hoVHS, where intra-VHS scattering should dominate\cite{hoVHS1}, and both superconductivity and pseudogap are absent.  We have proposed that this situation can arise for bilayer and trilayer cuprates, where the bonding band is heavily doped, has a large $|t'|$ value, and has less to do with superconductivity\cite{hoVHS1}.  Figure~\ref{fig:A3}(a) shows how a camelsback feature develops in the dispersion near $(\pi,0)$ when $t'$ extends beyond the critical value for the hoVHS.  This causes a pocket to split away from the main Fermi surface near $(\pi,0)$ [and $(0,\pi)$], blue Fermi surfaces in frames (b) and (c).  In turn, as doping is reduced away from thr camelsback range the Fermi soufaces become very flat near the antinodal points (green curves), leading to exceptionally strong nesting.  Such a  CDW instability may have already been observed.\cite{reent}

\section{Charge order at large doping}
\subsection{Nesting related origin of $S_2$ curve}
In Fig.~4(b) of the main text, we compare the $Q$ vector of the $S_2$ branch to a variety of models of Fermi surface nesting for Bi2201.  We contrast two tight-binding models, the first an accurate multi-parameter fit to experimental data\cite{JennySci} (the violet dot-dashed and dashed curves), the second to a three-parameter member of the cuprate `reference family'\cite{MBMB} used in Paper 1\cite{Paper1} (blue dashed curve) to simultaneously model all cuprates using a single tuning parameter for each cuprate family.  The NM nesting vector was taken as $Q_x=2k_{Fx}$, for the Fermi surface points $(-k_{Fx},\pi)$ and $(k_{Fx},\pi)$, leading to strong disagreement with the curve $S_2$.  We note however that the nesting $Q_x$ extrapolates to 0 at $x_{VHS}$, close to the linear extrapolation of the experimental $S_2$ curve, as predicted in the NPS model.  These results are consistent with a CDW calculation\cite{Gutzcharge}, where a $6\times 6$ crossed CDW calculation produced an electron Fermi surface analogous to the proposed QO surface, but the $4\times 4$ calculation could not produce nesting for any hole doping.  This also strengthens the case that the second phase in the NPS is a form of CDW.

\begin{figure}
\rotatebox{0}{\scalebox{0.4}{\includegraphics{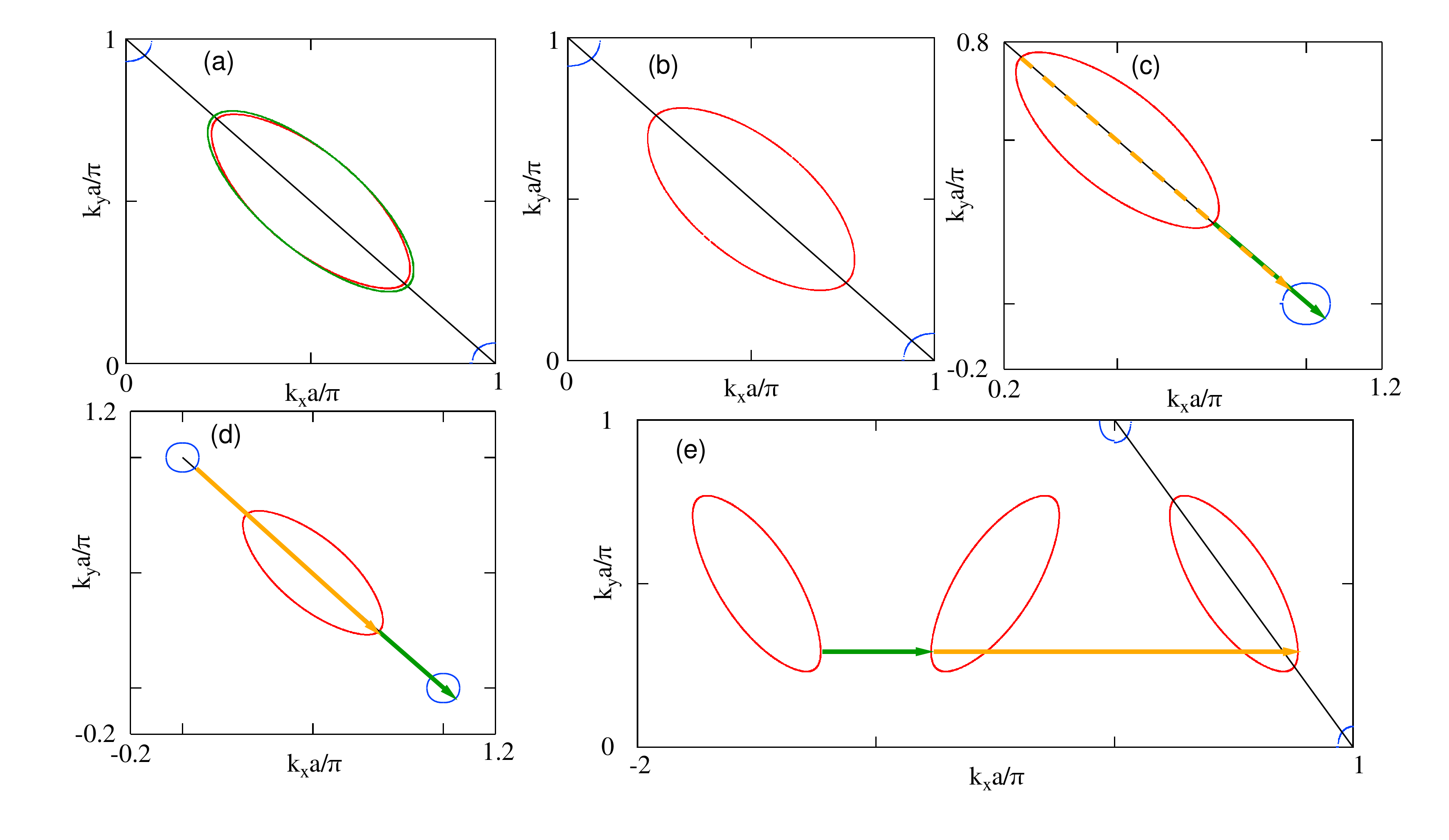}}}
\vskip0.5cm
\caption{
(a) AFM Fermi surfaces, using tight binding model of Ref.~\onlinecite{JennySci} (red and blue lines) or $3-t$ model with $t'/t=-0.2$ from Ref.~\onlinecite{Paper1} (green line), with AFM gap $\Delta=150$~meV.  Fermi surfaces are electron Fermi surface from upper magnetic band (blue) and hole Fermi surface from lower magnetic band (red or green).  (b) $3-t$ model for $t'/t=-0.26$.  (c) Electron-hole nesting, with arrows denoting nesting vectors $Q'$ (green) and $S'$ (orange).  The corresponding vectors for hot-spot nesting are shown in Fig.~3(c) of the Main Text.  (d,e) Illustrating $Q+S$ for electron-hole nesting (d) and hot-spot nesting (e).
}
\label{fig:A2}
\end{figure}

Recognizing the failure of models nesting the NM Fermi surface, it was proposed that nesting related to the AFM Fermi surface -- e.g., hot-spot\cite{EHud2} or Fermi arc\cite{Comin} nesting -- might prove a better starting point.  Here, we study nesting on the same AFM Fermi surface nesting model used in Paper 1\cite{Paper1}, comparing the two tight-binding models noted above, Fig.~\ref{fig:A2}(a,b).  We note the resemblance to the AFM pockets in electron-doped cuprates. While there are a number of candidates for nesting, we need a $Q$-vector that starts out large at low doping and gets smaller with increased doping.  The main candidates are shown by the green lines in Figs.~\ref{fig:A2}(c) and 3(c) of the Main Text.  The electron-hole nesting in frame (c) would seem to be excluded, as the nesting vector is diagonal, whereas we need horizontal and vertical $Q$s.  However, we allow for the possibility of an exotic two-dimensional CDW, where the $x$ and $y$ nesting vectors are $Q_x$ and $Q_y$, respectively.  The two narrow dashed lines in Fig.4(b) represent the two tight binding models, and it is clear that neither is a good match to the data.  However, Fig. 3(c), the equivalent of hot-spot\cite{EHud2} or Fermi arc\cite{Comin} nesting, provides a good fit, thick violet dashed line in Fig. 4(b).  In Figs.~3(d,e) we illustrate the star=shaped electron pockets expected for double nesting at high magnetic fields.

In Figs.~\ref{fig:A2}(c) and 3(c), there is a second nesting vector, $S$ (orange arrows).   Putting both arrows on the same line, Figs.~\ref{fig:A2}(d,e), it can be seen that their sum connects a point in one Brillouin zone to the equivalent point in a different Brillouin zone, leading to the relations $S'_x+Q'_x=\pi/a$ for frame (d) and $S_x+Q_x=2\pi/a$ for frame (e).

Finally, we comment on the two tight binding models, the N-component (Fig.~\ref{fig:A2}(a) and thick violet curve in Fig.~3(b)) and the 3-component (green curve in Fig.~\ref{fig:A2}(a) and thick light-blue curve in Fig.~3(b)).  The three component model is a member of the reference family used in Paper 1 to study the evolution of properties with cuprate families.  While it has a very similar shape of the hole Fermi surface, the electron pockets are only present for a small doping range, so the interpocket nesting is absent.  For a slightly larger $t'$ (Fig.~\ref{fig:A2}(b)) the electron pockets reappear, and we used this dispersion to plot interpocket nesting (thin light-blue curve in Fig.~3(b)).  Note that both tight-binding models provide good fits to the hot-spot nesting, Fig.~3b.
